\documentclass[12pt]{article}
\usepackage{amssymb,amsmath}

\usepackage{curves}
\usepackage{epic}
\usepackage{eepic}
\usepackage{epsfig}
\usepackage{verbatim}

\newcommand{\dd}{\mbox{\rm d}}
\newcommand{\wg}{\wedge}
\newcommand{\gam}{\gamma}

\newcommand{\dg}{\dagger}
\newcommand{\ddg}{\ddagger}
\newcommand{\tl}{\tilde}
\newcommand{\ul}{\underline}

\newcommand{\nnn}{\noindent}

\newcommand{\p}{\partial}

\newcommand{\be}{\begin{equation}}
\newcommand{\bear}{\begin{eqnarray}}

\newcommand{\ear}{\end{eqnarray}}
\newcommand{\ee}{\end{equation}}
\newcommand{\lbl}{\label}

\newcommand{\ci}{\cite}

\newcommand{\vs}{\vspace}
\newcommand{\hs}{\hspace}

\begin{document}

\begin{center}

\
\vs{1cm}

\baselineskip .7cm

{\bf \Large Quantum Field Theories in Spaces with Neutral Signatures} \\

\vs{2mm}

\baselineskip .5cm
Matej Pav\v si\v c

Jo\v zef Stefan Institute, Jamova 39, SI-1000, Ljubljana, Slovenia; 

email: matej.pavsic@ijs.si

\vs{3mm}

{\bf Abstract}
\end{center}

\baselineskip .43cm
{\small We point out that  quantum field theories based on the concept of Clifford
space and Clifford algebra valued-fields involve both positive and negative
energies. This is a consequence of the indefinite signature $(p,q)$ of  the Clifford
space. When the signature is neutral, $p=q$, then vacuum energy vanishes and there
is no cosmological constant problem. A question of the stability of such theories in
the presence of interactions arises. We investigate a toy model of the harmonic
oscillator in the space $M_{1,1}$. We have found that  in the presence of certain
interactions the amplitude of oscillations can remain finite. In general this is not
the case and the amplitude grows to infinity, but only when the two frequencies
are exactly the same. When they are even slightly different, the amplitude remains
finite and  the system is stable. We show how such oscillator comes from the
Stueckelberg action in curved space, and how it can be generalized to
field theories.}

\baselineskip .55cm

\section{Introduction}

Research in quantum gravity leads us to higher derivative theories that
imply negative energies. They have turned out to be a stumbling block
against further progress, because they imply vacuum instability.
Negative energies also occur in spaces with non Lorentzian signature,
$M_{p,q}$. Therefore, such spaces are commonly considered as unsuitable
for physics. In the literature it is usually stated that negative signatures
imply negative probabilities or ghosts.
In ref.\,(\ci{Woodard}--\ci{PavsicSaasFee}) it
has been pointed out that by an appropriate redefinition of vacuum,
instead of negative probabilities and positive energies,
one has positive probabilities and negative energies.

In this paper we investigate the issue of negative energies in some detail.
We consider a classical interacting oscillator in the space $M_{1,1}$.
Solutions of its equations of motion can be obtained numerically. We have
found that the solutions are not always runaway, unstable ones. In many
cases they have oscillatory behavior, and do not escape into infinity.
The same is true for a truncated quantum oscillator.

Then we consider field theories in field spaces with neutral signature,
an example of which is the Clifford space, discussed in Sec.\,2. The
vacuum energy of such fields is zero, which resolves the notorious
cosmological constant problem. But there remain the problem of
``instantaneous" vacuum decay of an interacting field. We show in
Sec.\,6  how this problem can be circumvented, and the ``explosion" of
the vacuum stabilized.

\section{Clifford space: An extension of spacetime}

An extended object, ${O}$, can be be sampled by a finite set of
parameters, for instance, by the center of mass coordinates, and by the
orientation of its axes of symmetries. Higher multipole deformations,
such as the dipole and the quadrupole ones, can also be taken into account.
For practical reasons, only a finite number of mulipoles can be taken into
account. Instead of the infinite number of degrees of freedom, we
consider only a finite number of degrees of freedom. We thus perform
a mapping from an infinite dimensional configuration space, associated
with the object ${O}$, to a finite dimensional subspace.

Extended objects of particular interest for theoretical physics are
strings and branes. They can be described by coordinate functions
$X^\mu (\xi^a)$, $\mu=0,1,2,...,N-1$, $a=0,1,2,...,n-1$, where
$n\le N$. Such a description is infinite dimensional. In refs.\,\ci{Aurilia}
it was pointed out how one can employ a finite description in terms of
a quenched mini superspace.

The idea has been further developed\,\ci{Castro}--\ci{PavsicMaxwelBrane}
by means of Clifford
algebras, a very useful tool for description of geometry\,\ci{Hestenes}.
Here we are interested in description of spacetime, $M_N$, and the objects
embedded in $M_N$.Therefore, let us start by considering the line element
in $M_N$:
\be
     Q = \dd s^2 = g_{\mu \nu} \dd x^\mu \dd x^\nu~, 
          ~~~~~~\mu,\nu=0,1,2,...N-1.
\lbl{2.1}
\ee
If we take the square root, $\sqrt{Q}$, we have the following possibilities:

\baselineskip 0.7cm

~~~~~\ i) $\sqrt{Q} = \sqrt{g_{\mu \nu} \dd x^\mu \dd x^\nu}$ ~~~~~~scalar

~~~~~ii)  $\sqrt{Q} = \gam_\mu \dd x^\mu$ ~~~~~~~~~~~~~~~vector

\baselineskip 0.55cm

\nnn Here $\gam_\mu$ are generators of the Clifford algebra $Cl(p,q)$,
$p+q=N$, satisfying
\be
    \gam_\mu \cdot \gam_\nu \equiv \frac{1}{2}(\gam_\mu \gam_\nu +
     \gam_\nu \gam_\mu ) = g_{\mu \nu} ,
\lbl{2.2}
\ee
where $g_{\mu \nu}$ is the metric of $M_N$.

The generators $\gam_\mu$ have the role of basis vectors of the spacetime
$M_N$. The symmetric product $\gam_\mu \cdot \gam_\nu$ represents
{\it the inner product}. The antisymmetric (wedge) product of two basis
vectors gives a unit bivector:
\be
    \gam_\mu \wg \gam_\nu \equiv \frac{1}{2}(\gam_\mu \gam_\nu -
     \gam_\nu \gam_\mu )
\lbl{2.2a}
\ee
and has thus the role of outer product. In analogous way we obtain
3-vectors, 4-vectors, etc..

We assume that the signature of an $N$-dimensional spacetime
is $(1,N-1)$, i.e., $(+ - - - ...)$. In the case of the 4-dimensional
spacetime we thus have the signature $(1,3)$, i.e., $(+ - - -)$.
The corresponding Clifford algebra is $Cl(1,3)$.

The basis of $Cl(1,N-1)$ is
\be
      \lbrace 1, \gam_\mu, \gam_{\mu_1} \wg \gam_{\mu_2},...,
       \gam_{\mu_1} \wg \gam_{\mu_2} \wg ... \wg\gam_{\mu_N} \rbrace .
\lbl{2.3}
\ee
A generic element, $X \in Cl(1,N-1)$, is a superposition
\be
    X=\sum_{r=0}^N \frac{1}{r!} X^{\mu_1 \mu_2 ...\mu_r}
    \gam_{\mu_1} \wg \gam_{\mu_2}\wg ... \wg\gam_{\mu_r} \equiv X^M \gam_M ,
\lbl{2.4}
\ee
called a {\it Clifford aggregate} or {\it polyvector}.

In refs. \ci{PavsicArena,PavsicMaxwelBrane,PavsicTachyonLocal}
it has been demonstrated that
$r$-vectors $X^{\mu_1 \mu_2 ...\mu_r}$ can be associated with closed
instantonic $(r-1)$-branes or open instantonic $r$-branes. A generic polyvector,
$X=X^M \gam_M$, can be associated with a conglomerate of (instantonic)
$r$-branes for various values of $r=0,1,2,...,N$.

Our objects are instantonic $r$-branes, which means that they are localized
in spacetime\footnote{The usual $p$-branes are localized in space,
but they are infinitely extended into a time-like direction, so that they
are $(p+1)$-dimensional worldsheets in spacetime.}.
They generalize  the concept of `event', a spacetime point,
$x^\mu,~\mu=0,1,2,3$. Instead of an event, we have now an
extended event, ${\cal E}$,  described by coordinates 
$X^{\mu_1 \mu_2 ...\mu_r}$,
$r=0,1,2,3,4$. The space of extended events is called {\it Clifford space},
$C$. It is a manifold whose tangent space at any of its points is
a Clifford algebra $Cl(1,3)$. If $C$ is a flat space, then it is isomorphic
to the Clifford algebra $Cl(1,3)$ with elements $X=\sum_{r=0}^4
\frac{1}{4!} X^{\mu_1 \mu_2 ...\mu_r}
    \gam_{\mu_1 \mu_2 ...\mu_r}\equiv X^M \gam_M$.

{\it In flat} $C$-space, the basis vectors are equal to the wedge product
\be
    \gam_M = \gam_{\mu_1} \wg \gam_{\mu_2}\wg ... \wg \gam_{\mu_r}
\lbl{2.5}
\ee
at every point ${\cal E} \in C$. This not true in {curved} $C$-space:
if we (parallel) transport a polyvector $A=A^M \gam_M$ from a point
${\cal E} \in C$ along a closed path back to the original point, ${\cal E}$,
then the orientation of the polyvector $A$ after such transport will not
coincide with the initial orientation of $A$. After the transport along
a closed path we will obtain a new polyvector $A'=A'^M \gam_M$. If,
in particular, the initial polyvector is one of the Clifford
algebra basis elements, $A=\gam_M$, i.e., an object with definite grade,
then the final polyvector will be $A'=A'^M \gam_M$, which is an object with
mixed grade. A consequence is that in curved Clifford space $C$,
basis vectors cannot have definite grade at all points of $C$.

The situation in a curved Clifford space, $C$, is analogous to that in a
usual curved space, where after the (parallel) transport along a closed
path, a vector changes its orientation. In Clifford space, a change of
orientation in general implies a change of a polyvector's grade, so
that, e.g., a definite grade polyvector changes into a mixed grade polyvector.

However, if we impose a condition that, under parallel transport, the grade
of a polyvector does not change, then one has a very special kind of
curved Clifford space\,\ci{CastroCurved}. In such a space, afer a parallel
transport along a closed path, the vector part $\langle A \rangle_1 =
a^\mu \gam_\mu$ changes into $\langle A' \rangle_1 =a'^\mu \gam_\mu$,
the bivector part $\langle A \rangle_2 =a^{\mu \nu} \gam_\mu\wg \gam_\nu$
changes into $\langle A' \rangle_2 =a'^{\mu \nu} \gam_\mu\wg \gam_\nu$,
etc., but one grade does not change into another grade. Such special
Clifford space, in which the consequences of curvature manifest themselves
{\it within} each of the subspaces with definite grade separately, but not
{\it between} those subspaces, is very complicated. We will not consider
such special Clifford spaces, because they are analogous to the usual
curved spaces of the product form $M= M_1 \times M_2 \times...M_n$,
where $M_i \subset M$ is a curved lower dimensional subspace of $M$, and
where only those (parallel) transports are allowed that bring tangent
vectors of $M_i$ into another tangent vectors of the same subspace
$M_i$.

The squared line element in Clifford space, $C$, is
\be
   \dd S^2 = G_{MN} \dd x^M \dd x^N = \d X^\ddg * \dd X 
   = \langle \dd X^\ddg \dd X \rangle .
\lbl{2.5a}
\ee
Here $\dd X = \dd x^M \gam_M$, and $\dd X^\ddg = \dd x^M \gam_M^\ddg$,
where $\ddg$ denotes the operation of inversion:
$(\gam_{\mu_1} \gam_{\mu_2} ... \gam_{\mu_r})^\ddg
 = \gam_{\mu_r}\gam_{\mu_{r-1}} ...\gam_{\mu_r}$. The metric of $C$ is
\be
    G_{MN} = \gam_M^\ddg * \gam_N = \langle \gam^\ddg \gam_N \rangle_0 ,
\lbl{2.6}
\ee
where $\langle ~~\rangle_0$ means the scalar part. A Clifford space with
such a metric has signature\,\ci{PavsicMaxwelBrane}
$(8,8)$, i.e., $(++++++++--------)$. This is ultrahyperbolic space with
neutral signature.

In the following sections we will show that, contrary to the wide spread
belief, the physics in spaces with signature $(n,n)$ makes sense.

\section{A toy model: Harmonic oscillator in the pseudo Euclidean space
$M_{r,s}$}.

\subsection{Case $M_{1,1}$}

In ref.\,\ci{PavsicPseudoHarm} we considered the harmonic oscillator
described by the Lagrangian
\be
    L = \frac{1}{2}(\dot x^2  - \dot y^2 ) 
      - \frac{1}{2}\omega ^2 (x^2  - y^2 ) .
\lbl{3.1}
\ee
The change of sign in front of the $y$-terms has no influence on the
equations of motion. A difference occurs when we calculate the
canonical momenta
\be
      p_x  = \frac{{\partial L}}{{\partial \dot x}} = \dot x\:,\qquad p_y  
      = \frac{{\partial L}}{{\partial \dot y}} =  - \dot y
\lbl{3.2}
\ee
The Hamiltonian
\be
     H = p_x \dot x + p_y \dot y - L = \frac{1}{2}(p_x^2  - p_y^2 ) 
      + \frac{{\omega ^2 }}{2}(x^2  - y^2 )
\lbl{3.3}
\ee
can have positive or negative values, but this does not mean that the
system is unstable. Namely, the equations of motion 
\be
    {\ddot x} = - \frac{\p V}{\p x}~,~~~~~~~
    {\ddot y} =  \frac{\p V}{\p y}~,~~~~
    V = \frac{{\omega ^2 }}{2}(x^2  - y^2 )
\lbl{3.4}
\ee
have different signs in front of the force terms.
The criterion for stability of the $y$-degree of freedom is that the
potential has to have a {\it maximum} in the $(y,V)$-plane\,\ci{PavsicSaasFee}.

\setlength{\unitlength}{.8mm}

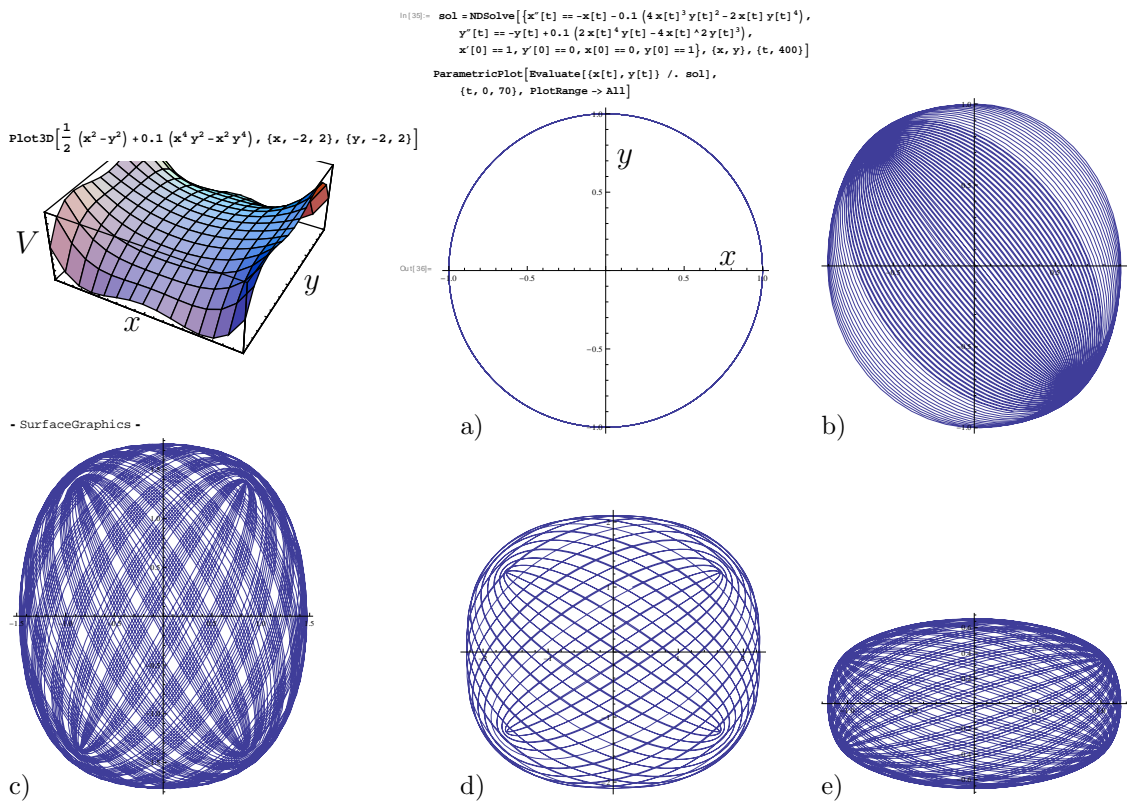
\begin{figure}[h!]
\hs{3mm} \begin{picture}(120,126)(25,0)
\put(25,60){\includegraphics[scale=0.5]{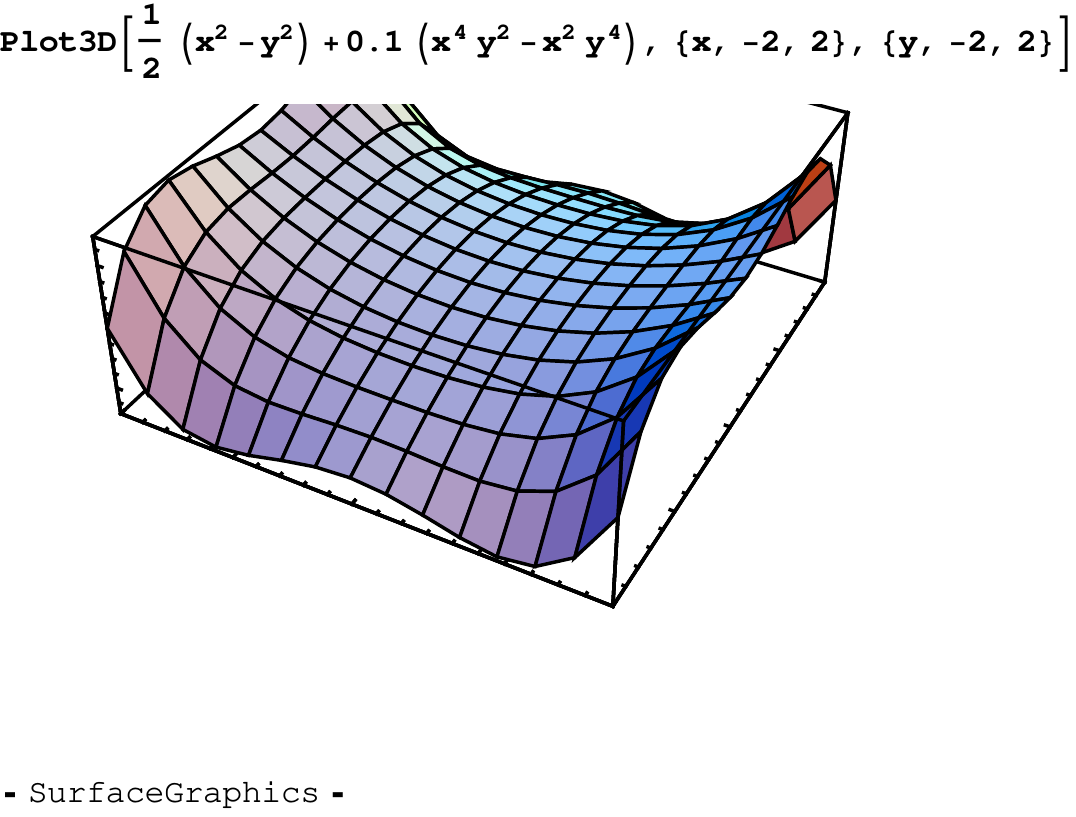}}
\put(90,60){\includegraphics[scale=0.43]{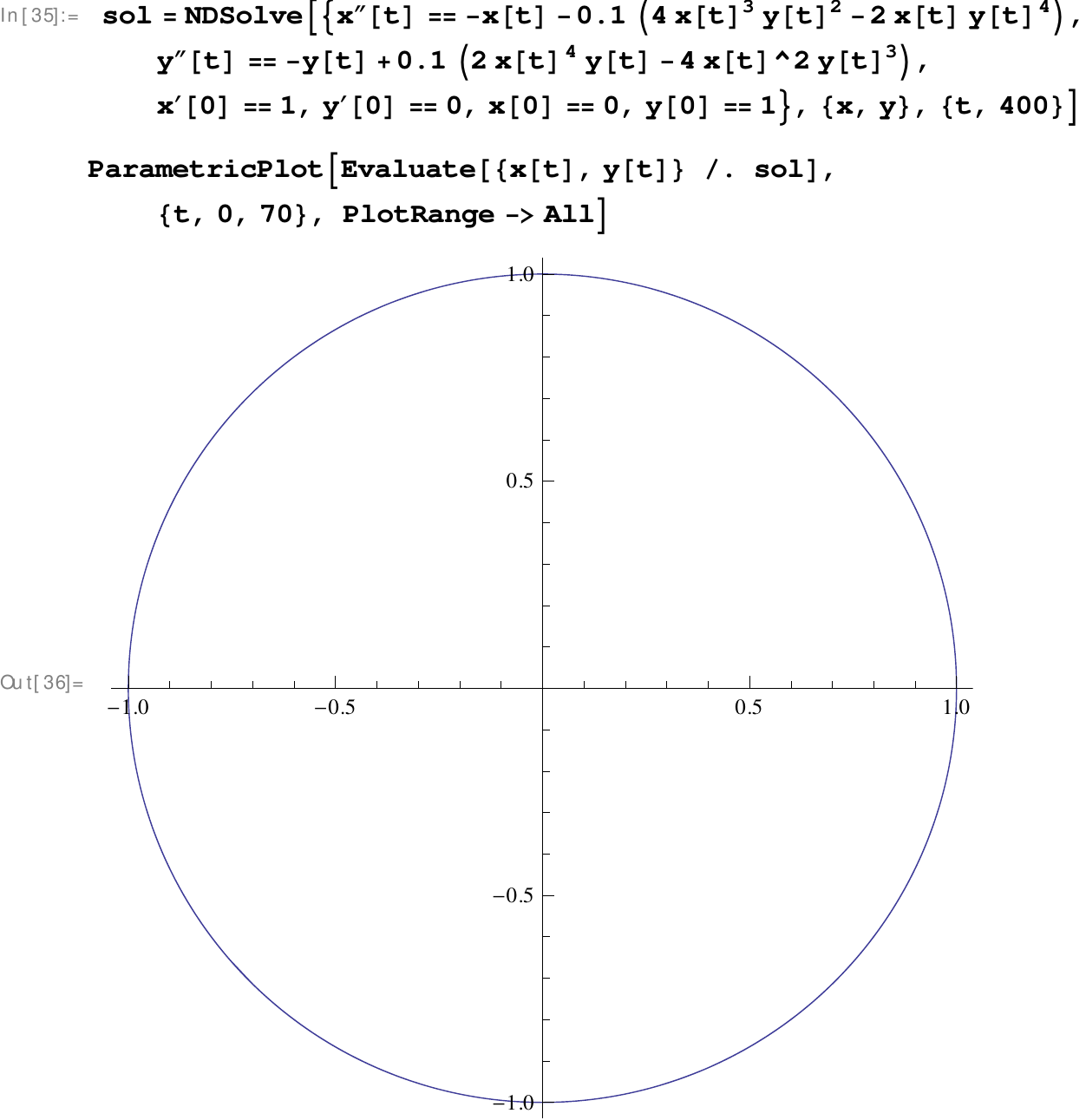}}
\put(160,60){\includegraphics[scale=0.4]{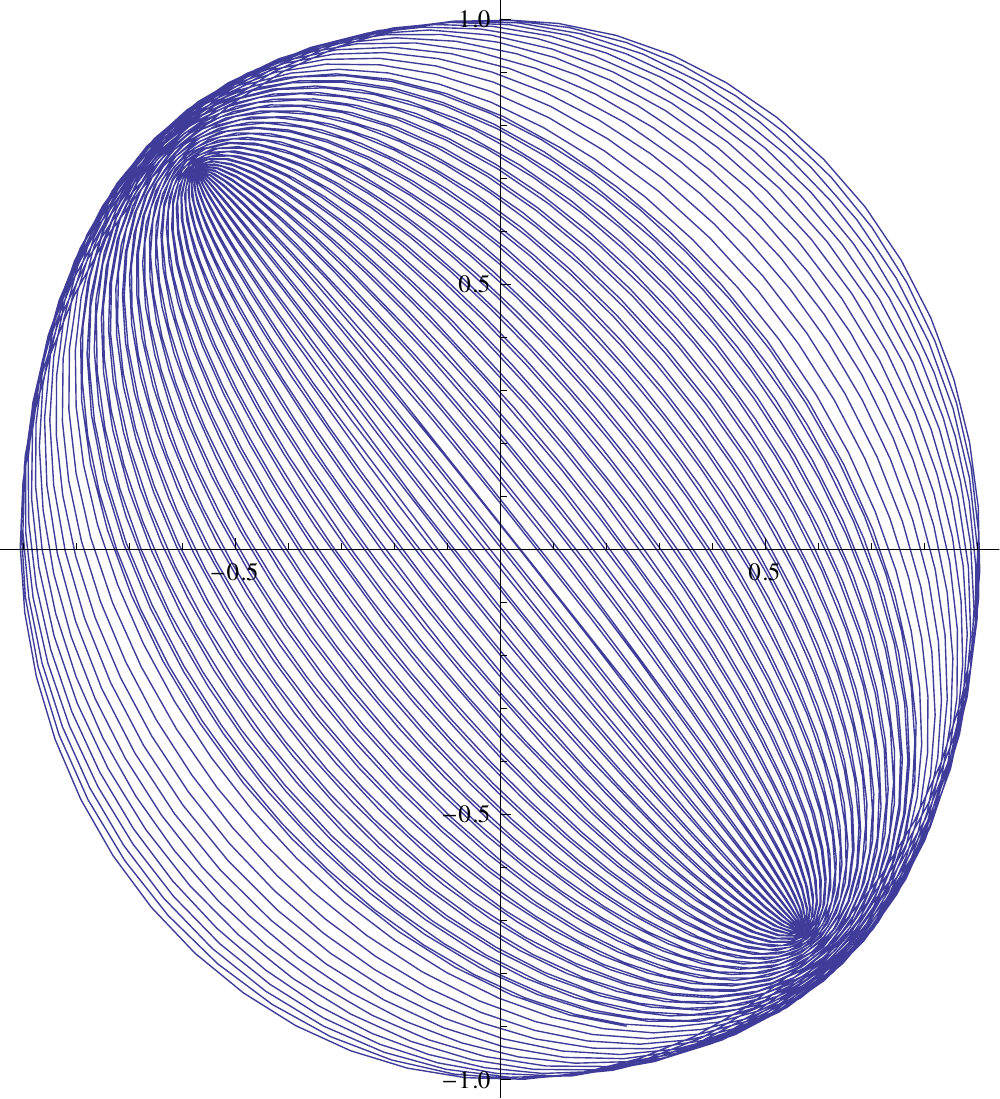}}
\put(25,0){\includegraphics[scale=0.4]{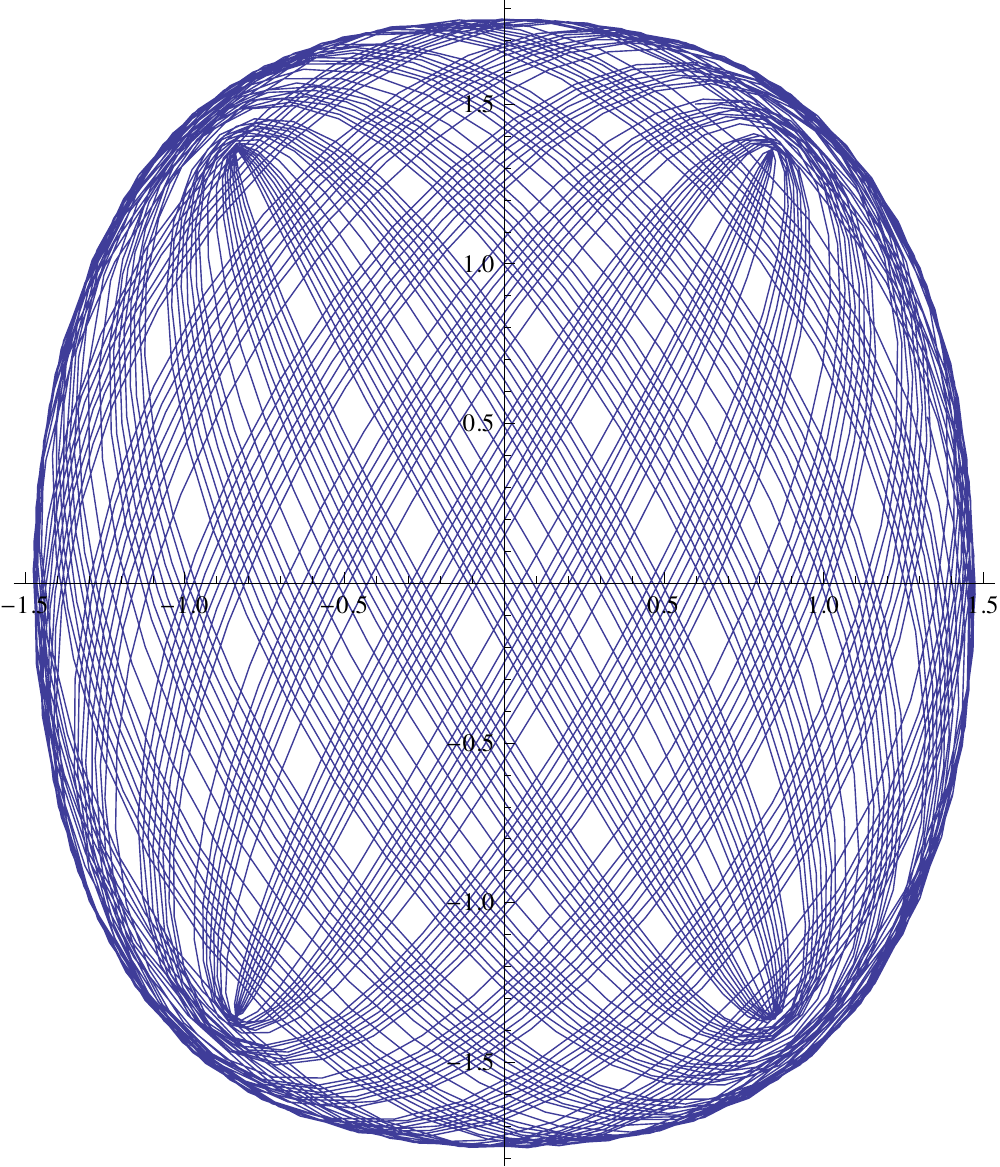}}
\put(100,0){\includegraphics[scale=0.4]{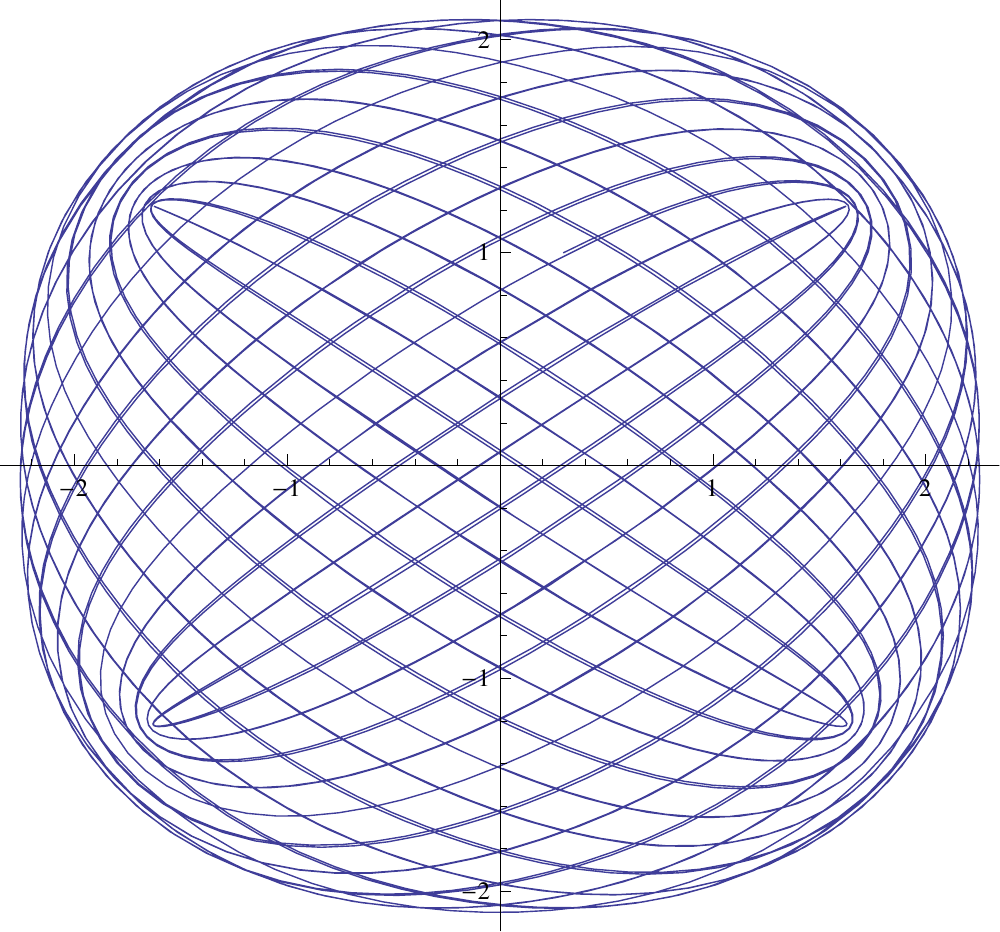}}
\put(160,0){\includegraphics[scale=0.4]{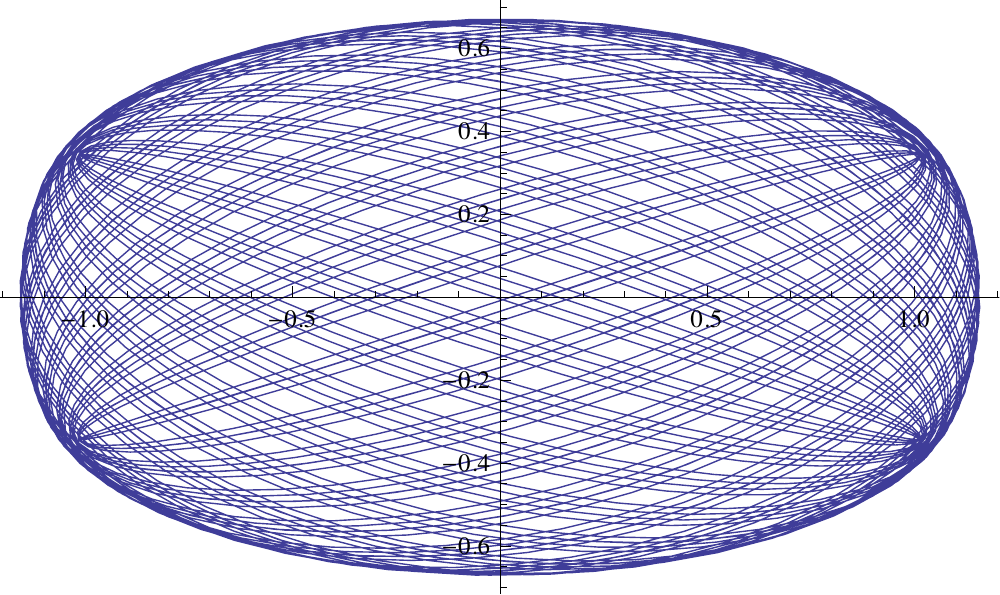}}

\put(26,90){$V$}
\put(44,77){$x$}
\put(74,84){$y$}

\put(143,88){$x$}
\put(126,105){$y$}

\put(100,60){\footnotesize a)}
\put(160,60){\footnotesize b)}
\put(25,0){\footnotesize c)}
\put(100,0){\footnotesize d)}
\put(160,0){\footnotesize e)}

\end{picture}

\caption{\footnotesize Examples of solutions, obtained by Mathematica,
for various initial conditions,
of the interactive pseudo-Euclidean oscillator for the potential
$V=(1/2)(x^2-y^2)+ V_1$, with $V_1=0.1 (x^4 y^2-x^2 y^4)$.
a) ${\dot x}(0)=1,~{\dot y}(0)=0$, $x(0)=0$, $y(0)=1$; 
b) ${\dot x}(0)=0.9,~{\dot y}(0)=0$, $x(0)=0$, $y(0)=1$;
c) ${\dot x}(0)=0.9,~{\dot y}(0)=0.2$, $x(0)=0.6$, $y(0)=1.5$;
d) ${\dot x}(0)=2,~{\dot y}(0)=1$, $x(0)=0.3$, $y(0)=1$;
e) ${\dot x}(0)=0.2,~{\dot y}(0)=0.2$, $x(0)=0.3$, $y(0)=1$.}

\end{figure} 

Stability could be destroyed if we include an extra interaction term
into the potential $V$. But for
\be
    V = \frac{1}{2}(x^2  - y^2 ) + \lambda (x^4 y^2 - x^2 y^4)
\lbl{3.5}
\ee
we find, by solving numerically the equations of motion, that the system
is stable (Fig.\,1). We find the same phenomenon also for other potentials (Fig.\,2).
Those figures show that the trajectories in the $(x,y)$
space remain confined, and do not escape into infinity, 
which reveals that the interacting pseudo-Euclidean
oscillator is not automatically unstable. Later we will study this
in more detail.

\begin{figure}[h!]
\hs{3mm} \begin{picture}(120,80)(25,0)
\put(25,0){\includegraphics[scale=0.4]{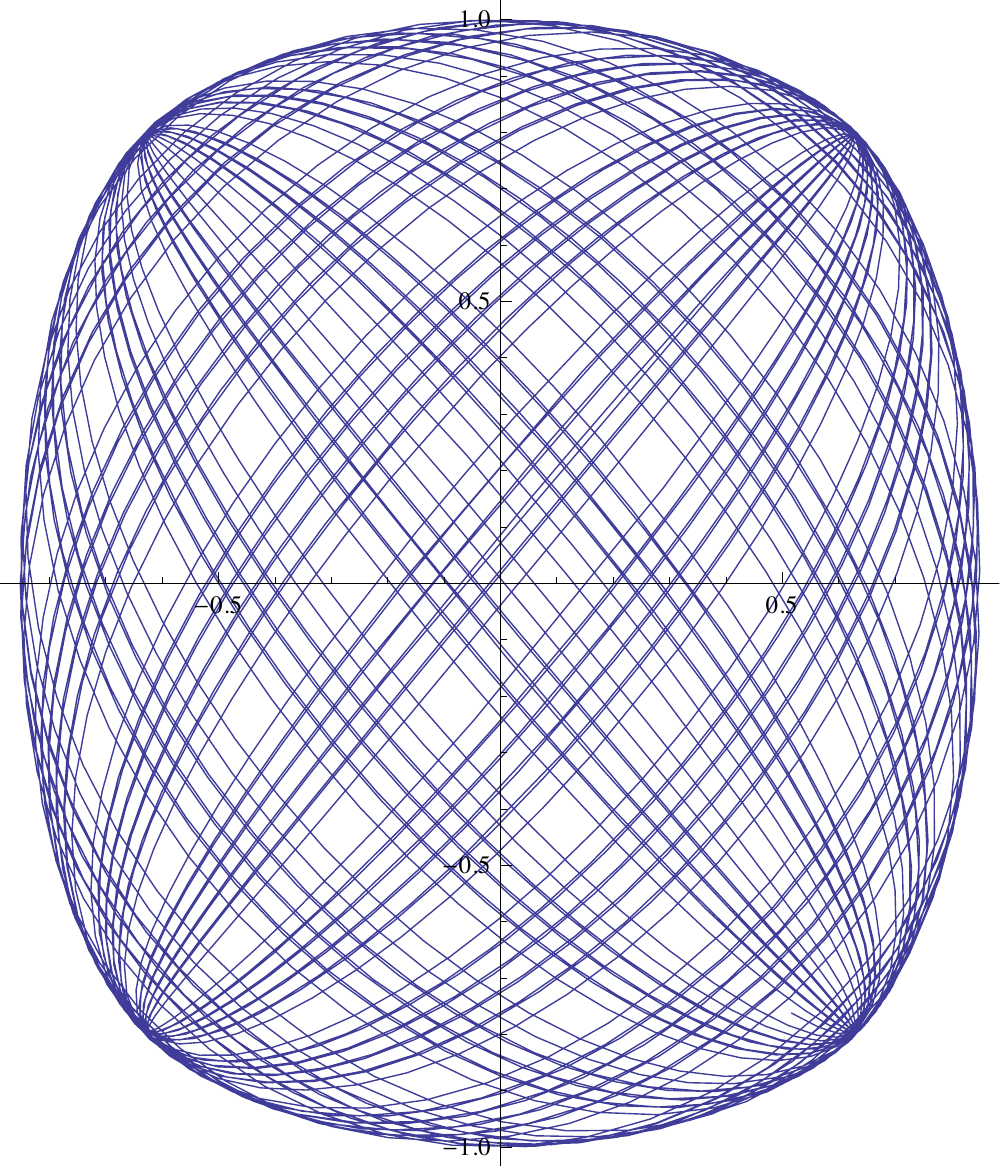}}
\put(90,0){\includegraphics[scale=0.45]{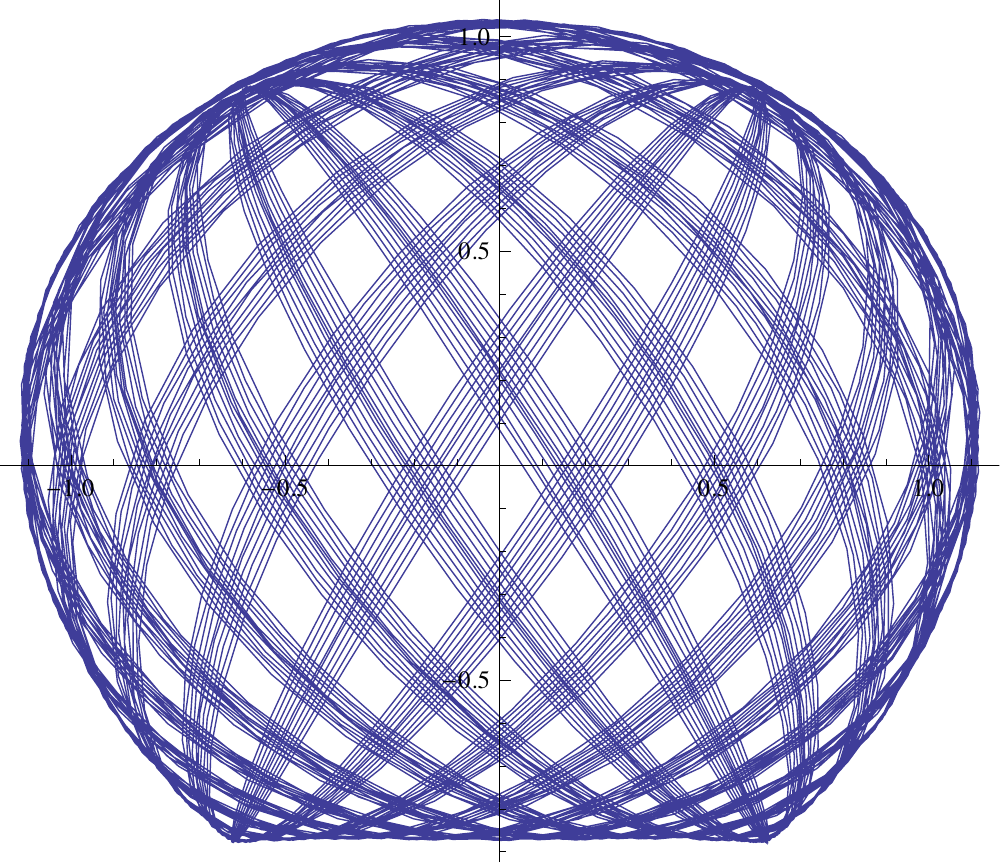}}
\put(160,0){\includegraphics[scale=0.4]{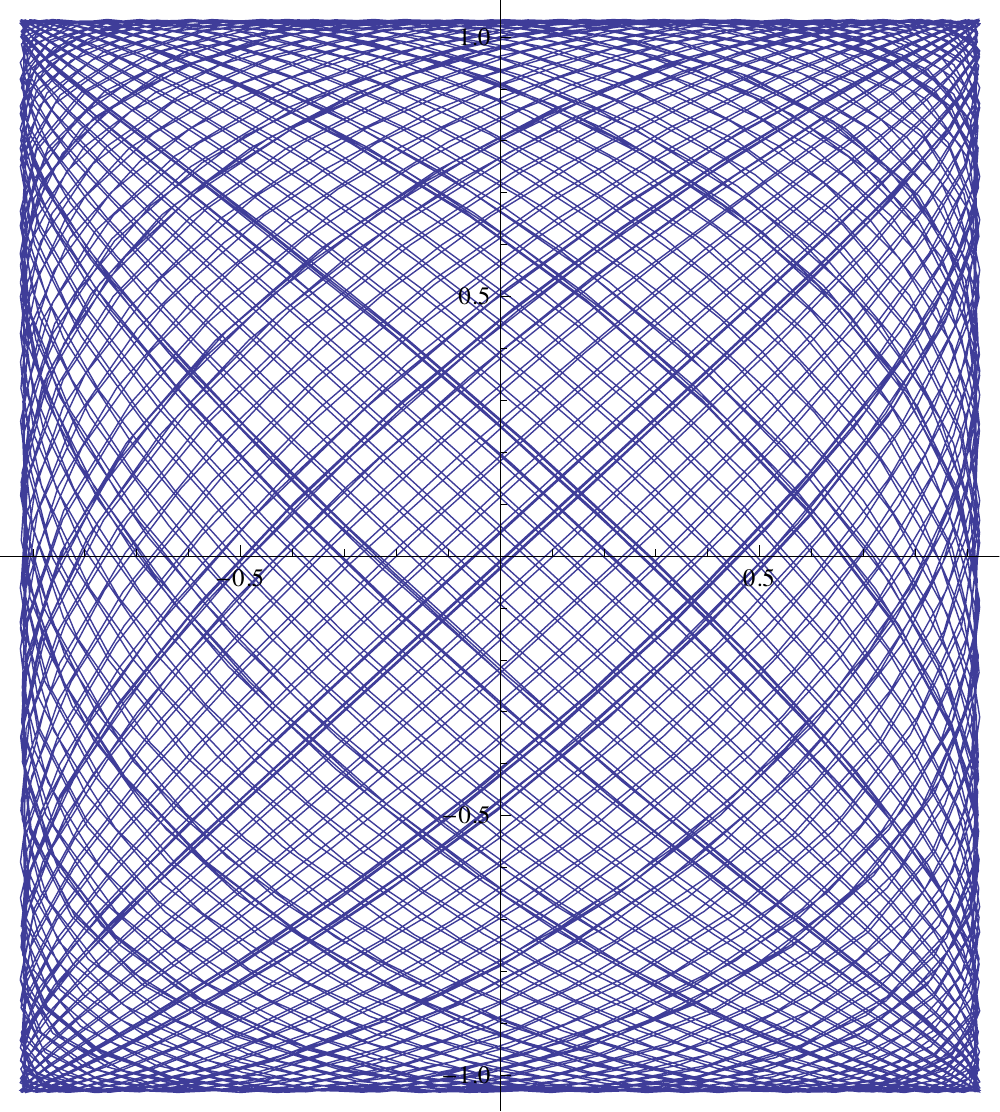}}

\put(32,76){$^{V_1=0.1 (x^4 y^2+x^2 y^4)}$}
\put(32,70){$^{{\dot x}(0)=0.9,~{\dot y}(0)=0}$}
\put(32,65){$^{x(0)=0,~y(0)=1}$}

\put(98,76){$^{V_1=0.1 (x^4 y-x^3)}$}
\put(98,70){$^{{\dot x}(0)=0.8,~{\dot y}(0)=0.2}$}
\put(98,65){$^{x(0)=0.2,~y(0)=0.9}$}

\put(165,76){$^{V_1=0.1 (x^4 + y^4)}$}
\put(165,70){$^{{\dot x}(0)=1,~{\dot y}(0)=0.2}$}
\put(165,65){$^{x(0)=0,~y(0)=1}$}

\put(149,21){$x$}
\put(117,51){$y$}

\end{picture}

\caption{\footnotesize Examples of solutions, for various initial conditions
and  potentials, of the interactive pseudo-Euclidean oscillator}
\end{figure}
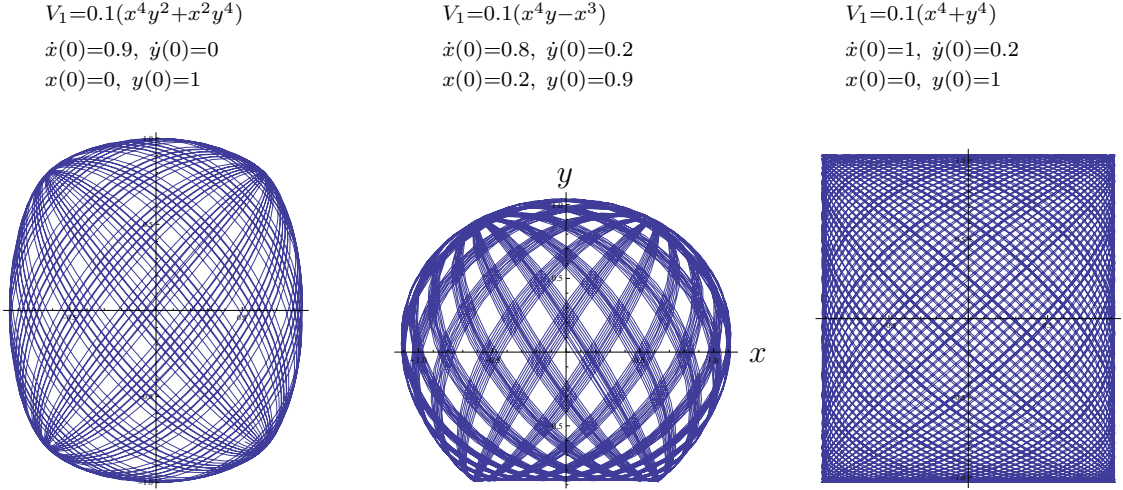

In refs.\,\ci{PavsicPseudoHarm,PavsicSaasFee} we also studied the quantized non interacting system
(\ref{3.1}) by replacing the coordinates and momenta with the operators,
satisfying $[x,p_x]=i$, $[y,p_y]=i$, $[x,y]=[p_x,p_y]=0$, and by introducing
the operators
\bear
 &&c_x  = \frac{1}{{\sqrt {2} }}(\sqrt \omega  \,x + 
\frac{i}{{\sqrt \omega  }}\,p_x )\:,\quad c_x^\dg   
= \frac{1}{{\sqrt 2 }}(\sqrt {\omega}  \,x 
- \frac{i}{{\sqrt \omega  }}\,p_x ) \lbl{3.6} \\ 
 &&c_y  = \frac{1}{{\sqrt {2} }}(\sqrt {\omega}  \,y 
+ \frac{i}{{\sqrt {\omega}  }}\,p_y )\:,\quad c_y^\dg   
= \frac{1}{{\sqrt {2} }}(\sqrt {\omega}  \,y 
- \frac{i}{{\sqrt {\omega}  }}\,p_y ) \lbl{3.7}
 \ear
satisfying the commutation relations 
\be
 [c_x ,c_x^\dg  ] = 1\:,\qquad [c_y ,c_y^\dg  ] = 1, \lbl{3.8}\ee 
\be
  [c_x ,c_y ] = [c_x^\dg  ,c_y^\dg  ] = 0 \lbl{3.9} 
\ee
The Hamilton operator then reads
\be
      H = \omega \,(c_x^\dg c_x + c_x c_x^\dg - c_y^\dg  c_y- c_y c_y^\dg ).
\lbl{3.10}
\ee
Let us define vacuum so that it is annihilated by $c_x$ and $c_y$:
\be
     c_x |0\rangle = 0~,~~~~~~~c_y |0\rangle =0.
\lbl{3.11}
\ee

If we arrange the terms in the Hamiltonian so that the creation operators
are on the left, we obtain
\be
      H = \omega \,(c_x^\dg  c_x  - c_y^\dg  c_y )
\lbl{3.12}
\ee
which has vanishing vacuum expectation value,
\be
     \langle 0|H|0\rangle = 0.
\lbl{3.13}
\ee
The excited states $c_x^\dg |0\rangle$, $c_x^\dg c_x^\dg|0\rangle$,...,
have positive energies, whereas the states $c_y^\dg |0\rangle$,
$c_y^\dg c_y^\dg|0\rangle$,..., have negative energies. But all those
states have positive norms, because the commutators
(\ref{3.8}) are positive.

In the Schr\"odinger representation, $x,y \in \mathbb{R}$, $p_x=-i \p/\p x$,
$p_y=-i \p/\p y$, and $\langle x,y|0 \rangle \equiv \psi_0 (x,y)$.
The relations (\ref{3.11}) become
\bear
 &&\frac{1}{2}\left( {\sqrt \omega  x + \frac{1}{{\sqrt \omega  }}\,
\frac{\partial }{{\partial x}}} \right)\psi _0 (x,y) = 0 \lbl{3.14}\\ 
 &&\frac{1}{2}\left( {\sqrt \omega  y + \frac{1}{{\sqrt \omega  }}\,
\frac{\partial }{{\partial y}}} \right)\psi _0 (x,y) = 0 \lbl{3.15} 
\ear
and their solution is the following vacuum wave function:
\be
     \psi_0 = A {\rm e}^{\frac{1}{2}\omega (x^2 + y^2)}  .
\lbl{3.16}
\ee
It is normalized according to $\int \psi_0^2 \, \dd x \dd y = 1$
if $A=\sqrt{\omega/\pi}$, and it satisfies the Schr\"odinger equation
$H \psi = E \psi$, with $E=0$.

\section{Generalization to $M_{r,s}$}

The system (\ref{3.1}) can be generalized\,\ci{PavsicPseudoHarm} to a space
of arbitrary dimension and signature, $M_{r,s}$. The Lagrangian is then
\be
    L = \frac{1}{2}\dot x^a \dot x_a  - \frac{1}{2}\omega ^2 x^a x_a .
\lbl{3.17}
\ee
The corresponding Hamiltonian is
\be
      \frac{1}{2} p^a p_a  + \frac{1}{2}\omega ^2 x^a x_a ,
\lbl{3.18}
\ee
where $p_a  = \partial L/\partial \dot x^a  
= \dot x_a  = \eta _{ab} \dot x^b$ are the canonical momenta,
and $\eta_{ab}= {\rm diag} (1,1,1,...,-1,-1,-1)$ the metric tensor
with signature $(r,s)$. Upon quantization, the classical variables
$x^a,~p_a$ are replaced by the operators, satisfying
\be
    [x^a ,p_b ] = i\delta ^a _b \quad {\rm{or}}\quad [x^a ,p^b ] = i\eta ^{ab}
\lbl{3.19}
\ee
Because the momenta $p_a$ are canonically conjugated to $x^a$, a natural choice
of creation and annihilation operators is
\bear
     &&c^a  = \frac{1}{{\sqrt 2 }}\left( {\sqrt \omega  x^a  + 
    \frac{i}{{\sqrt \omega  \,}}p_a } \right) \lbl{3.20}\\ 
     &&{c^a}^\dg   = \frac{1}{{\sqrt 2 }}\left( {\sqrt \omega  x^a  
    - \frac{i}{{\sqrt \omega  \,}}p_a } \right) \lbl{3.21} 
\ear
the commutation relations being
\be
   [c^a,{c^b}^\dg]=\delta^{ab}~,~~~~~[c^a,c^b]=[{c^a}^\dg,{c^b}^\dg]=0.
\lbl{3.22}
\ee
The Hamiltonian then becomes
\be
    H = \frac{1}{2}\,\omega \,(c_a^\dg  c^a  + c^a c_a^\dg  ) .
\lbl{3.23}
\ee
Defining vacuum according to
\be
     c^a |0 \rangle = 0,
\lbl{3.24}
\ee
and using $c^a c_a^\dg   = \eta _{ab} c^a c^{b\dg }$
$= \eta _{ab} (c^{b\dg } c^a  + \delta ^{ab} ) = c^{a\dg } c_a  + r - s$,
the Hamiltonian (\ref{3.23}) can be rewritten as
\be
     H = \omega \,(c_a^\dg  c^a  + \frac{r}{2} - \frac{s}{2}) .
\lbl{3.25}
\ee
Vacuum expectation of the latter Hamiltonian is zero if the signature
is neutral, i.e., if $r=s$.

In eqs.\,(\ref{3.20}),(\ref{3.21}) we composed the creation and annihilation
operators in terms of the {\it contravariant} components $x^a$ and the
{\it covariant} components $p_a$. But in the literature, the following operators
are used:
\bear
 &&a^a  = \frac{1}{2}\left( {\sqrt \omega  \,x^a  
+ \frac{i}{{\sqrt \omega  }}\,p^a } \right) \lbl{3.26}\\ 
 &&{a^a}^\dg   = \frac{1}{2}\left( {\sqrt \omega  \,x^a  
- \frac{i}{{\sqrt \omega  }}\,p^a } \right) \lbl{3.27} 
\ear
They are composed in terms of $x^a$ and $p^a = \eta^{ab} p_b$, and satisfy
\be
    [a^a,a_b^\dg]={\delta^a}_b~, ~~~~~[a^a,{a^b}^\dg]= \eta^{ab}.
\lbl{3.28}
\ee
The Hamiltonian is
\be
     H = \frac{1}{2}\omega \,(a^{a\dg } a_a  + a_a a^{a\dg } ).
\lbl{3.29}
\ee

There are two possible definitions of the vacuum:

~~ {\it Possibility} I
\be
     a^a |0 \rangle = 0.
\lbl{3.30}
\ee
This is the usual definition. Rewriting the Hamiltonian in the form
\be
    H = \omega \,\,(a^{a\dg } a_a  + \frac{r}{2} + \frac{s}{2}),
\lbl{3.31}
\ee
we see that the vacuum energy, $\langle 0|H|0 \rangle =(r+s)/2$, is positive.
The eigenvalues of $H$ are all positive. But there exist negative norm states,
called ghosts.

~~ {\it Possibility} II
\be
    a^{\bar a} |0\rangle  = 0~,~~~~~ {a^{\ul a}}^\dg |0\rangle  = 0 ,
\lbl{3.32}    
\ee    
where we have split the index $a$ according to
\be
    {\bar a}=1,2,...,r~;~~~~~{\ul a} = r+1,r+2,...,r+s .
\lbl{3.33}
\ee
Such definition of vacuum (and its consequences) has been used
in refs.\,\ci{Jackiw, Woodard,PavsicPseudoHarm}. Writing the Hamiltonian
(\ref{3.29}) in the form
\be
    H = \omega \,\,(a^{\bar a\dg } a_{\bar a}  
    + a_{\ul a} {a^{\ul a}}^\dg  + \frac{r}{2} - \frac{s}{2}) ,
\lbl{3.34}
\ee
we see that the energy of the vacuum (\ref{3.32}) is $E=\omega (r-s)/2$..
If $r=s$, then the vacuum energy vanishes. The excitation states have positive
or negative energies, $E=\omega (m-n +r/2 -s/2)$, depending on the mode
$(m,n)$ of excitation. There are no negative norm (ghost) states.

The vacuum definition (\ref{3.32}) is equivalent to the definition (\ref{3.24}), because
the operators $a^a$, $a^{a\dg}$ are just rewritten operators $c^a$ and $c^{a \dg}$
From the definition (\ref{3.20}), (\ref{3.21}) of the latter operators it follows that the vacuum
defined according to (\ref{3.32}) is covariant under the transformations of the
group SO(r,s).

\section{Non interacting quantum field theory}

\subsection{The scalar fields}

In refs.\,\ci{PavsicPseudoHarm} we considered the scalar field theory described
by the action
\be
I[\varphi ^a ] = \frac{1}{2}\int {\dd^4 } x\,\sqrt { - g} 
(g^{\mu \nu } \partial _\mu  \varphi ^a \,\partial _\nu  \varphi ^b  
- m^2 \varphi ^a \varphi ^b )\gamma _{ab} ,
\lbl{4.1}
\ee
where $\gam_{ab}$ is a metric in the space of fields $\varphi^a (x)$ at a
fixed point $x\equiv x^\mu$ of spacetime. We will consider the case
$\gam_{ab} = {\rm diag} (1,1,1,...,-1,-1,-1) = \gam^{ab}$ and
$g_{\mu \nu} = {\rm diag} (1,-1,-1,-1) = g^{\mu \nu}$.

The canonical momenta are
\be
\pi _a  = \frac{{\partial L}}{{\partial \partial _0 \varphi ^a }} 
= \partial ^0 \varphi _a  = \partial _0 \varphi _a  \equiv \dot \varphi _a ,
\lbl{4.2}
\ee
and the Hamiltonian is
\be
 H = \frac{1}{2}\int {\dd^3 } x\,(\dot \varphi ^a \dot \varphi ^b  
 - \partial _i \varphi ^a \partial ^i \varphi ^b  
 + m^2 \varphi ^a \varphi ^b )\gamma _{ab}  .
\lbl{4.3}
\ee
In the quantized theory, $\varphi^a(x)$ and $\pi_a (x)$ are operators
satisfying
\be
  [\varphi ^a ({\bf{x}}),\pi _b ({\bf{x}}')] 
  = i\delta ^3 ({\bf{x}} - {\bf{x}}'){\delta^a}_b  .
\lbl{4.4}
\ee
Expanding the field according to
\be
   \varphi ^a  = \int {\frac{{{\rm{d}}^3 {\bf{k}}}}{{(2\pi )^3 }}} 
   \,\frac{1}{{2\omega _{\bf{k}} }}(a^a ({\bf{k}}){\rm{e}}^{ - ikx} 
    + {a^a}^\dg  ({\bf{k}}){\rm{e}}^{ikx} ) ,
\lbl{4.5}
\ee
where $\omega_{\bf k} = \sqrt{m^2+{\bf k}^2}$, and
\be
    [a^a ({\bf{k}}),a_b^\dg  ({\bf{k'}})] 
    = (2\pi )^3 \,2\omega _{\bf{k}} \,\delta ^3 
    ({\bf{k}} - {\bf{k}}') {\delta^a}_b ,
\lbl{4.6}
\ee
i.e.,
\be
   [a^a ({\bf{k}}),{a^b}^\dg  ({\bf{k'}})] 
   = (2\pi )^3 \,2\omega _{\bf{k}} \,\delta ^3 
   ({\bf{k}} - {\bf{k}}')\gamma ^{ab} ,
\lbl{4.7}
\ee
the Hamiltonian (\ref{4.3}) becomes
\be
   H = \frac{1}{2}\int {\frac{{\dd^3 {\bf{k}}}}{{(2\pi )^3 }}} 
   \,\frac{{\omega _{\bf{k}} }}{{2\omega _{\bf{k}} }}({a^a}^\dg  
   ({\bf{k}})a^b ({\bf{k}}) + a^a ({\bf{k}}){a^b}^\dg 
    ({\bf{k}}))\gamma _{ab} .
\lbl{4.8}
\ee
Analogously to eq.\,(\ref{3.32}), we define the vacuum as
\be
   a^{\bar a} ({\bf{k}})|0\rangle  = 0
   \;,\quad \,\,\,{a^{\ul a}}^\dg  ({\bf{k}})|0\rangle  = 0 ,
\lbl{4.9}
\ee
where we split the index $a$ into the part ${\bar a}$, runing over
the positive, and the part ${\ul a}$, runing over the negative
signature components.

Using (\ref{4.7}) we can rewrite the Hamiltonian (\ref{4.8}) into the
form
\be
    H = \int {\frac{{{\rm{d}}^3 {\bf{k}}}}{{(2\pi )^3 }}} 
    \frac{{\,\omega _{\bf{k}} }}{{2\omega _{\bf{k}} }}({a^{\bar a}}^\dg 
     ({\bf{k}})a_{\bar a} ({\bf{k}}) + a^{} ({\bf{k}})a_{} ^\dg  ) 
     + \frac{1}{2}\int {{\rm{ d}}^3 } {\bf{k}}\,\omega _{\bf{k}} 
     \delta ^3 (0)(r - s) .
\lbl{4.10}
\ee
If the signature has equal number of plus and minus signs, i.e., if
$r=s$, then the zero point energies cancel out from the Hamiltonian.

\subsection{Generalization to Clifford space}

\subsubsection{The generalized Klein-Gordon equation}

In ref.\,\ci{PavsicBook,PavsicKaluza,PavsicKaluzaLong,PavsicSpinorInverse} it has been proposed to consider the Clifford algebra
valued field in $C$-space:
\be
      \Phi(X) = \phi^A (X) \gam_A ,
\lbl{4.11}
\ee
where $\gam_A \equiv (1/r!) \gam_{a_1} \wg \gam_{a_2} \wg ...
\wg \gam_{a_p}$, $r=0,1,2,...,n$, is a basis of the Clifford algebra
$Cl(1,n-1)$ of the spacetime $M_{1,n-1}$.

In general, the field depends on coordinates $X^M\equiv X^{\mu_1 ...\mu_r}$
of $C$-space. In this paper we will consider the case in which the
field depends on spacetime coordinates only. The action is then
\be
    I = \frac{1}{2}\int {\dd^n } x\,\sqrt { - g} (g^{\mu \nu } \partial _\mu 
     \varphi ^A \partial _\nu  \varphi ^B  - m^2 \varphi ^A \varphi ^B )G_{AB} ,
\lbl{4.12}
\ee
where $\mu=0,1,2,...,n-1$. The metric is defined as
$G_{AB} = \gam_A^\ddg * \gam_B \equiv \langle \gam_A^\ddg \gam_B \rangle_0$,
where $\langle~~ \rangle_0$ means the scalar part of the expression, and
the operation $\ddg$ reverses the order of vectors. It turns
out\,\ci{PavsicSaasFee} that such metric has signature $(R,S)$ with
$R=S$, where $R+S=2^n$ is the dimension of $Cl(1,n-1)$. Because the signature
is neutral, the action (\ref{4.12}) is just like the action (\ref{4.1}).
Following the same procedure as before, we obtain that the vacuum energy
vanishes. Therefore, in such theory there is no cosmological constant
problem. Recall that in Einstein's equations there is a term with
the cosmological constant, $\lambda g_{\mu \nu}$, and the term
with the stress-energy tensor, $8 \pi G T_{\mu \nu}$. The vacuum of
a quantum field gives $T_{\mu \nu} = \rho g_{\mu \nu}$, where
$\rho$ is the energy density. The two contributions sum to an
effective cosmological constant $\Lambda  =  \lambda  + 8 \pi G \rho$.
Here $\lambda$ is a free parameter that can have in principle
any values. Usual field theoretic calculations give infinite vacuum
energy density $\rho$. When taking into account the cutoff at the Planck scale,
it turns out that $\rho$ is 120 orders of magnitude bigger than expected
from observations. Therefore, the effective cosmological constant, $\Lambda$,
is too big as well. This is the notorious cosmological constant problem.
Here we have found a way, how $\rho$ can vanish. The cosmological constant
problem is thus resolved, because what remains is $\Lambda = \lambda$,
and there is no longer the annoying huge contribution coming from
quantum vacuum. How $\lambda$ can be theoretically
calculated remains, of course, a problem. But this is
another sort of problem from the one, why the $\rho$ is so big.
The small observed cosmological constant could be a residual effect of
something else, e.g., of a spacetime filling brane\,\ci{Bandos,PavsicWheeler}.

\subsubsection{Generalized Dirac equation (Dirac-K\"ahler equation)}

Another possibility is to assume that the Clifford algebra valued field
(\ref{4.11}) satisfies the Dirac-K\"ahler equation\,\ci{DiracKaehler}
\be
   (i \gam^\mu \p_\mu - m)\Phi = 0.
\lbl{4.13}
\ee
Instead of expanding $\Phi$ in terms of $\gam_A = (1,\gam_{a_1},\gam_{a_1}
\wg \gam_{a_2},..., \gam_{a_1} \wg...\wg \gam_{a_n})$, we can expand it in terms
of the spinor basis, $\xi_{\tl A}$ of $Cl(1,n-1)$, considered in
refs.\,\ci{SpinorBasis,PavsicSpinorInverse,PavsicPhaseSpace}. Then
\be
   \Phi= \psi^{\tl A} \xi_{\tl A} \equiv \psi^{\alpha i} \xi_{\alpha i} ,
\lbl{4 14}
\ee
where $\alpha$ is the spinor index of a left minimal ideal of $Cl(1,n-1)$,
whereas $i$ runs over $2^{n/2}$ left ideal of $Cl(1,n-1)$. From now on
we will consider the case $n=4$.

The $\gam^\mu$ are abstract objects, the Clifford numbers, i.e., the
elements
of $Cl(1,3)$. They can be represented as matrices according to
\be
   \langle (\xi ^{\tilde A} )^\ddag  \gamma ^\mu  \xi _{\tilde B} \rangle _S  
   \equiv {(\gamma ^\mu  )^{\tilde A}}_{\tilde B} ,
\lbl{4.15}
\ee
where the subscript $S$ means the normalized scalar part of the expression
(see\,\ci{PavsicKaluzaLong}).
Multiplying eq.\,(\ref{4.13}) from the left by $(\xi^{\tl A})^\ddg$ and taking
the scalar part, we obtain
\be
   \left( {i\,{(\gamma ^\mu  )^{\tilde A}}_{\tilde B} \,\partial _\mu  
    - m{\delta^{\tilde A}}_{\tilde B} } \right)\psi ^{\tilde B}  = 0 .
\lbl{4.16}
\ee

The $16 \times 16$ matrices, ${(\gamma ^\mu  )^{\tilde A}}_{\tilde B}$, can
be factorized as ${(\gamma ^\mu  )^{\tilde A}}_{\tilde B}$
$  = {(\gamma ^\mu  )^\alpha }_\beta  {\delta ^i}_j $, and eq.\,(\ref{4.16})
written in the form
\be
    \left( {i\,{(\gamma ^\mu  )^\alpha} _\beta  \partial _\mu  
     - m{\delta ^\alpha }_\beta  } \right)\psi ^{\beta i}  = 0 ,
\lbl{4.17}
\ee
or shortly,
\be
    (i\,\gamma ^\mu  \partial _\mu   - m)\psi ^i  = 0 .
\lbl{4.18}
\ee
In the last equation, the spinor index, $\alpha$, has been  omitted,
and kept only the index $i=1,2,3,4$ denoting four left ideals. Bear in
mind that the $i$ as an index (superscript or subscript) has different a
role than the $i$ as a factor in front of $\gam^\mu$, in which case it is
the imaginary unit, $i^2 = -1$.

The action is
\be
I = \int {{\rm{d}}^4 x\,\,} \bar \psi ^i (i\,\gamma ^\mu  \partial _\mu   
- m)\psi ^j z_{ij} ,
\lbl{4.19}
\ee
where $z_{ij}$ is the metric in the space of ideals. It comes\footnote{
For more details see ref.\,\ci{PavsicKaluzaLong}.}
from the metric $z_{{\tl A}{\tl B}}$ of the 16-dimensional spinor space
of $Cl(1,3)$:
\be
  (\xi _{\tilde A} )^\ddag  *\xi _{\tilde B}  = z_{\tilde A\tilde B}  
  = z_{(\alpha i)(\beta j)}  = z_{\alpha \beta } z_{ij} ,
\lbl{4.20}
\ee
where
\be
\,z_{ij}  = \left( \begin{array}{l}
 1\,\,\,\,\,\,\,0\,\,\,\,\,\,\,\,0\,\,\,\,\,\,0 \\ 
 0\,\,\,\,\,\,\,1\,\,\,\,\,\,\,\,0\,\,\,\,\,\,0 \\ 
 0\,\,\,\,\,\,0\,\,\,\,\, - 1\,\,\,\,\,\,0 \\ 
 0\,\,\,\,\,\,0\,\,\,\,\,\,\,\,0\,\,\, - 1 \\ 
 \end{array} \right)\,\,,\,\,\,\,\,\,\,\,\,\,\,\,
 \,z_{\alpha \beta }  = \left( \begin{array}{l}
 1\,\,\,\,\,\,\,0\,\,\,\,\,\,\,\,0\,\,\,\,\,\,0 \\ 
 0\,\,\,\,\,\,\,1\,\,\,\,\,\,\,\,0\,\,\,\,\,\,0 \\ 
 0\,\,\,\,\,\,0\,\,\,\,\, - 1\,\,\,\,\,\,0 \\ 
 0\,\,\,\,\,\,0\,\,\,\,\,\,\,\,0\,\,\, - 1 \\ 
 \end{array} \right)  .
\lbl{4.21}
\ee

The fields $\psi^i \equiv \psi^{\alpha i} (x)$ and their conjugate momenta\footnote{
They come from the action (\ref{4.19}) written in the explicit form
$I= \int \dd^4 x \, \psi_{\tl A}^* \left ( {(\gam^\mu)^{\tl A}}_{\tl B} \, \p_\mu
- m \right ) \psi^{\tl B}$, from which we obtain $\pi_{\tl C} = \p {\cal L}/
\p {\dot \psi}^{\tl c}=i \psi_{\tl A}^* {(\gam^0)^{\tl A}}_{\tl C}$
 $=i \psi_{\alpha i}^* {(\gam^0)^\alpha}_\delta {\delta^i}_j$\ $=
i \psi_{\alpha j}^* {(\gam^0)^\alpha}_\delta =i \psi^{* \beta i} z_{\alpha \beta}
z_{ij} {(\gam^0)^\alpha}_\delta$. Because $z_{\alpha \beta} {(\gam^0)^\alpha}_\delta
=\delta_{\alpha \beta}$ (see ref.\,\ci{PavsicKaluzaLong}), we have
$\pi_{\delta j} = i {\psi^{* \delta}}_j$.}
$\pi_{\alpha i} (x)= i {\psi^{* \alpha}}_i (x)$ satisfy the equal time
anticommutation relations
\be
    \lbrace \psi^{\alpha i} (t, {\bf x}), \pi_{\beta j} (t,{\bf x'}) \rbrace ,
  = i {\delta^\alpha}_\beta {\delta^i}_j \delta^3 ({\bf x} -{\bf x'})
\lbl{4.21a}
\ee
\be
\lbrace \psi^{\alpha i} (t, {\bf x}), \psi^{\beta j} (t,{\bf x'}) \rbrace = 0
~,~~~~~~
  \lbrace \pi_{\alpha i}^\dg (t, {\bf x}), \pi_{\beta j}^\dg (t,{\bf x'})
    \rbrace = 0 .
\lbl{4.21b}
\ee

The Hamiltonian, belonging to the action (\ref{4.19}) is
\be
  H = \int {{\rm{d}}^3 x} \,\bar \psi ^i ( - i\gamma ^r \partial _r 
   + m)\psi ^j z_{ij} .
\lbl{4.22}
\ee
Expanding the $\psi^i$ in terms of the annihilation and creation operators,
we obtain
\be
  H = \sum\limits_{n = 1}^2 {\frac{{{\rm{d}}^3 {\bf p}}}{{(2\pi )^3 }}\,m\,
  \left( {b_s^{i\dg } ({\bf p})\,b_s^j ({\bf p})
   - d_s^i ({\bf p})\,d_s^{j\dg }({\bf p}) } \right)}
   \,z_{ij} .
\lbl{4.23}
\ee
The index $i=1,2,3,4$ distinguishes the spinors of different left
ideals of $Cl(1,3)$. The index $s=1,2$ is the usual one that distinguishes
`spin up' and `spin down' states.
We have the following anticommutation relations:
\be
   \lbrace {b_s}^i ({\bf p}),{{b_{s'}}^j}^\dg ({\bf p'})=
   (2 \pi)^3 \frac{E}{M}\, \delta^3 ({\bf p}-{\bf p'}) \delta_{s s'} z^{ij} ,
\lbl{4.23a}
\ee
\be
 \lbrace {d_s}^i ({\bf p}),{{d_{s'}}^j}^\dg ({\bf p'}) \rbrace=
   (2 \pi)^3 \frac{E}{M}\, \delta^3 ({\bf p}-{\bf p'}) \delta_{s s'} z^{ij} ,
\lbl{4.23b}
\ee
where $E=|\sqrt{{\bf p}^2+m^2}$.

Let us now split the index according to $i=({\bar i},{\ul i})$, where
${\bar i}=1,2$ and ${\ul i}=3,4$, and define vacuum as follows:
\be
   \begin{array}{l}
 b_s^{\bar i} |0\rangle  = 0\,,\,\,\,\,\,\,\,~d_s^{\bar i} |0\rangle  = 0\,\,\,\,\, \\ 
 {{b_s}^{\ul i}} |0\rangle  , 
 = 0\,,\,\,\,\,\,\,\,{d_s}^{\ul i} |0\rangle  = 0 . \\ 
 \end{array}
\lbl{4.24}
\ee
The terms in the Hamiltonian (\ref{4.23}) can be arranged so that we obtain
$$H= \sum_{s=1}^2 \frac{{\rm d}^3 p}{(2\pi )^3 }\,m\,
   \left \{  \left  [ {{b_s}^{\bar i}}^\dg ({\bf p}) {b_s}^{\bar j} ({\bf p})
   +{{d_s}^{\bar i}}^\dg ({\bf p}) {d_s}^{\bar j} ({\bf p}) \right ] 
   z_{{\bar i}\,{\bar j}} \right . $$
   $$\hs{3.3cm}+  \left . \left [ {{b_s}^{\ul i}}^\dg ({\bf p}) {{b_s}^{\ul j}} ({\bf p})
   +{{d_s}^{\ul i}}^\dg ({\bf p}) {{d_s}^{\bar j}} ({\bf p}) \right ] 
   z_{{\ul i}\,{\ul j}}  \right \} $$
\be   
    + \sum_{s=1}^2 \dd^3 {\bf p} \, E \delta (0)
       (-\delta^{{\bar i}\,{\bar j}} z_{{\bar i}\,{\bar j}}
       - \delta^{{\ul i}\,{\ul j}} z_{{\ul i}\,{\ul j}})
\lbl{4.25}
\ee
The term with $z_{{\bar i}{\bar j}}$ gives positive. whereas the term with
$z_{{\ul i}{\ul j}}$ gives negative expectation values. In the term with
$\delta (0)$, the positive and negative contributions cancel out,
because $z_{{\bar i} {\bar j}}= \delta_{{\bar i} {\bar j}}$, and
$z_{{\ul i} {\ul j}}= -\delta_{{\ul i} {\ul j}}$. Therefore,
the vacuum expectation value of this Hamiltonian is zero.

Each fermion $\psi^i$ couples to the corresponding gauge field. The Casimir
force between two metallic plates, consisting of $\psi^i$, $i=1$, is not
expected to vanish in this theory\,\ci{PavsicPseudoHarm}.
The vacuum expectation value $\langle H \rangle = \langle T^{00} \rangle$
is the source of the gravitational field. Because $\langle 0 |H|0 \rangle = 0$,
the cosmological constant vanishes. There is no problem of the huge
cosmological constant.

\section{Presence of interactions}

\subsection{Classical oscillator}

We will now add an interaction term to the Lagrangian for an oscillator
in the pseudo Euclidean space $M_{1,1}$:
\be
   L = \frac{1}{2}(\dot x^2  - \dot y^2 ) - V\,,\,\,\,\,\,\,\,\,\,\,\,
   V = \frac{\omega }{2}(x^2  - y^2 ) + V_1 \,\,
\lbl{5.1}
\ee
Equations of motions are
\bear
 &&\ddot x + \omega ^2 x + \frac{{\partial V_1 }}{{\partial x}} = 0, \lbl{5.2}\\ 
 &&\ddot y + \omega ^2 y - \frac{{\partial V_1 }}{{\partial y}} = 0, \lbl{5.3} 
\ear
As an example. we will study the following interaction:
\be
   V_1  = \frac{\lambda }{4}(x^2  - y^2 )^2 .
\lbl{5.4}
\ee
So we have
\bear
 &&\ddot x + \omega ^2 x + \lambda \,x(x^2  - y^2 ) = 0, \lbl{5.5} \\ 
 &&\ddot y + \omega ^2 y + \lambda \,y(x^2  - y^2 ) = 0. \lbl{5.6} 
\ear

\setlength{\unitlength}{.8mm}

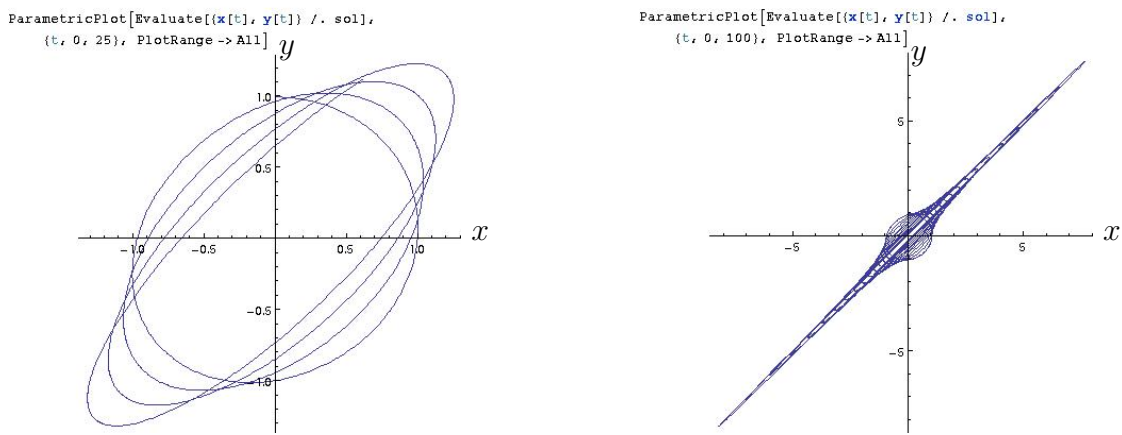
\begin{figure}[h!]
\hs{3mm} \begin{picture}(120,70)(25,0)
\put(25,0){\includegraphics[scale=0.4]{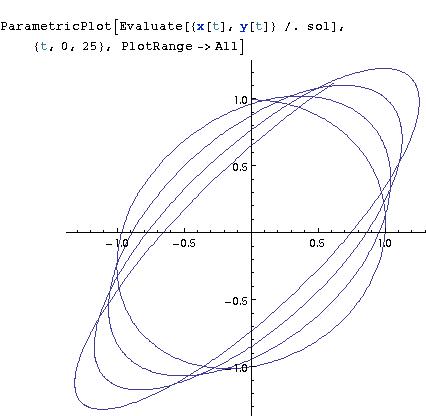}}
\put(130,0){\includegraphics[scale=0.4]{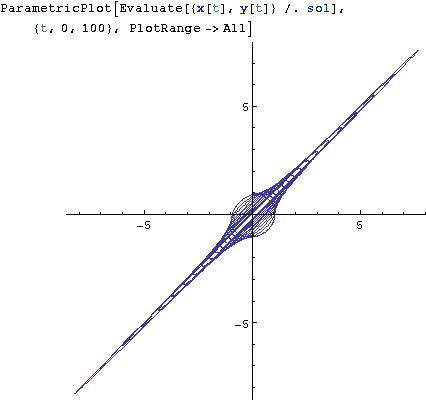}}

\put(102,32){$x$}
\put(70,63){$y$}

\put(207,32){$x$}
\put(175,62){$y$}

\end{picture}

\caption{\footnotesize The trajectory in the $(x,y)$ space of the system
described by eqs.\,(\ref{5.5}),(\ref{5.6}). Left: For the time
period from $t=0$ to $t=25$). Right: For the time period from
$t=0$ to $t=100$.}
\end{figure} 

These equations can be solved numerically by the program Mathematica.
In Figs. 3,4  we show the result for $\omega =1$, $\lambda =0.1$ and the
initial conditions ${\dot x} (0)=1$, ${\dot y} (0)=0$, $x(0)=0$, $y(0)=1$.
We see that the system is unstable in the sense that the amplitudes
of $x(t)$, ${\dot x}(t)$ grow to infinity. The total energy,
$E_{\rm tot}=({\dot x}^2 - {\dot y}^2)/2$ $+V(x,y)$, remains constant (Fig.4).

\setlength{\unitlength}{.8mm}
\begin{figure}[h!]

\hs{3mm} \begin{picture}(120,60)(25,0)

\put(15,0){\includegraphics[scale=0.5]{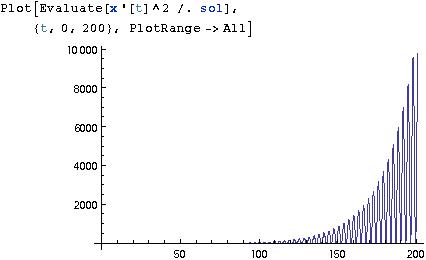}}
\put(120,0){\includegraphics[scale=0.4]{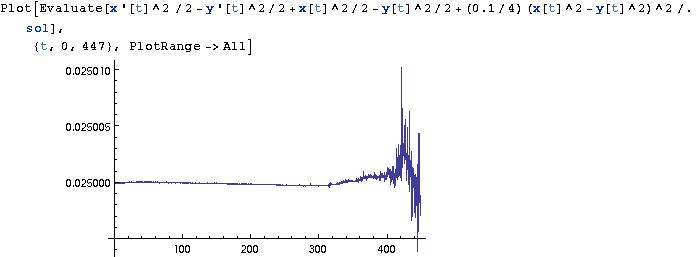}}

\put(39,42){$\frac{{\dot x}^2}{2}$}
\put(110,2){$t$}

\put(142,31){$E_{\rm tot}$}
\put(197,2){$t$}

\end{picture}

\caption{\footnotesize Left: The kinetic energy ${\dot x}^2/2$ as  function of
time. The envelop of oscillations grows to infinity. Right: The total
energy, $E_{\rm tor}$, remains constant within the numerical error. }
\end{figure}
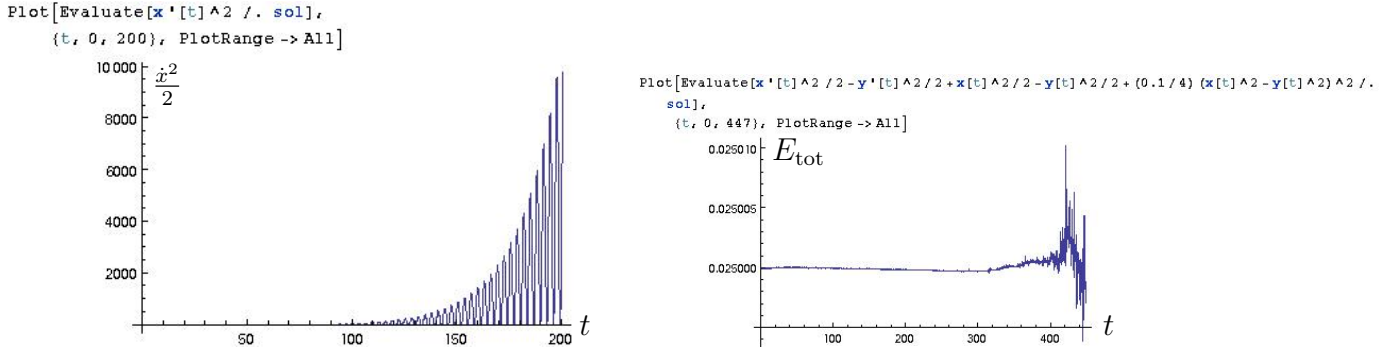

In Fig.\,5 we have the solution for the initial conditions
${\dot x} (0)=1$, ${\dot y} (0)=-1.2$, $x(0)=0$, $y(0)=0.5$. Here also the
system is unstable.

\setlength{\unitlength}{.8mm}

\begin{figure}[h!]
\hs{3mm} \begin{picture}(120,103)(25,0)
\put(20,10){\includegraphics[scale=0.45]{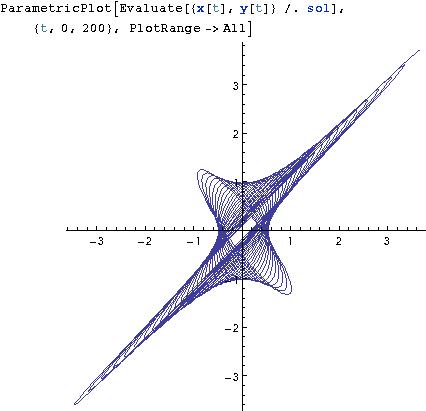}}
\put(120,60){\includegraphics[scale=0.4]{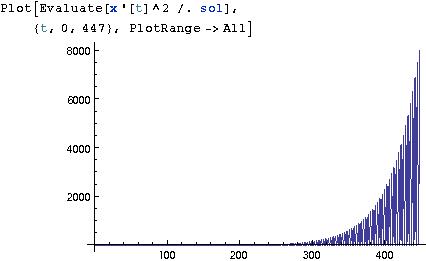}}
\put(120,0){\includegraphics[scale=0.4]{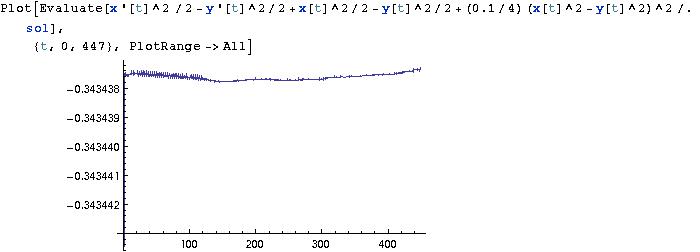}}

\put(102,42){$x$}
\put(70,81){$y$}

\put(138,92){$\frac{{\dot x}^2}{2}$}
\put(197,62){$t$}

\put(142,25){$E_{\rm tot}$}
\put(197,2){$t$}

\end{picture}

\caption{\footnotesize The solution of the system
described by eqs.\, (\ref{5.5}),(\ref{5.6}), for the initial conditions
${\dot x} (0)=1$, ${\dot y} (0)=-1.2$, $x(0)=0$, $y(0)=0.5$.}
\end{figure}
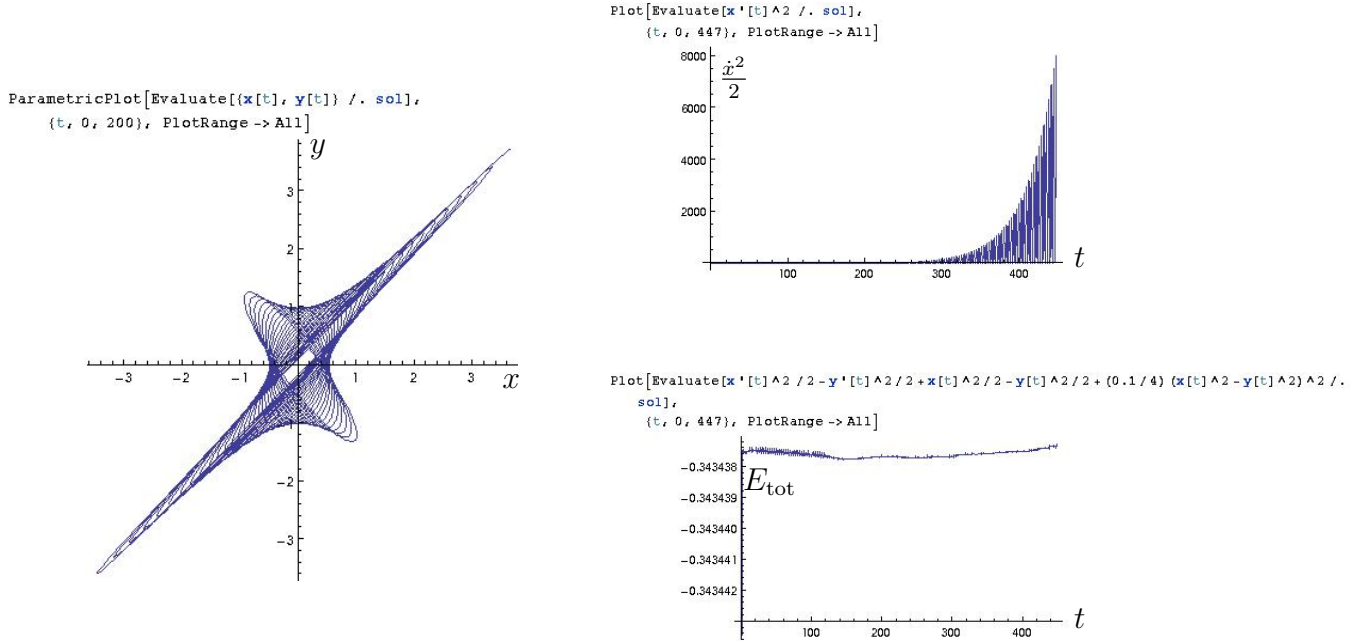

Something fascinating happens, if instead of eqs.\,(\ref{5.5}),(\ref{5.6}),
we take slightly different equations,
\bear
 &&\ddot x + \mu [\omega ^2 x + \lambda \,x(x^2  - y^2 )] = 0, \lbl{5.7} \\ 
 &&\ddot y + \nu [\omega ^2 y + \lambda \,y(x^2  - y^2 )] = 0. \lbl{5.8}
\ear

\setlength{\unitlength}{.8mm}

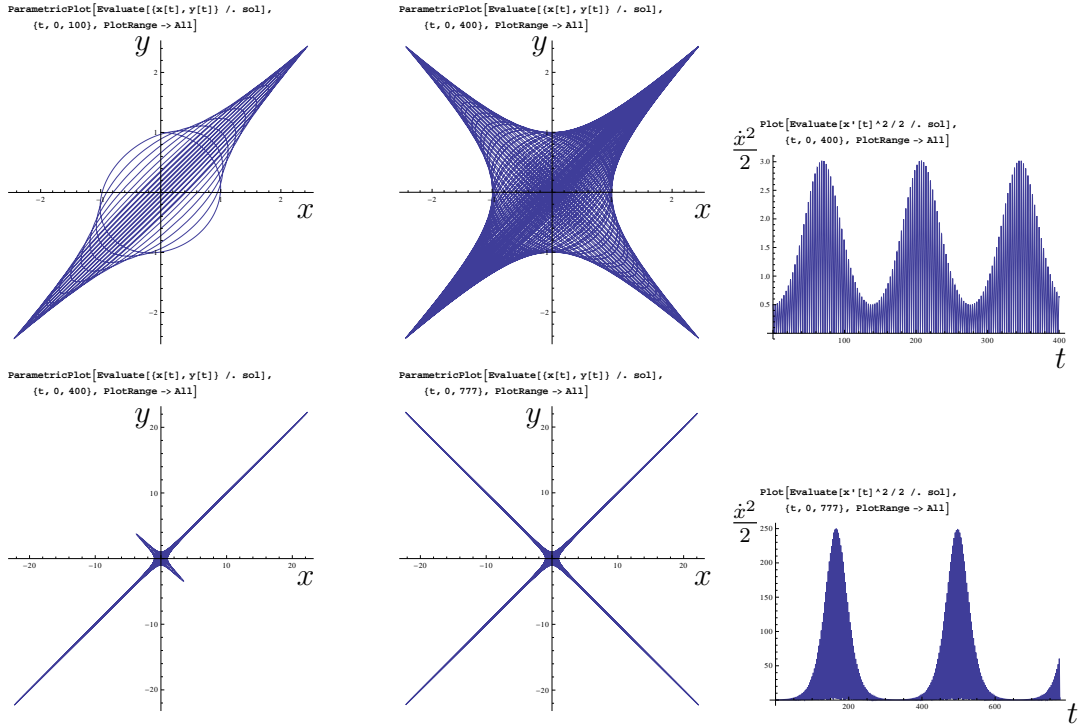
\begin{figure}[h!]
\hs{3mm}
\begin{picture}(120,100)(25,0)

\put(25,61){\includegraphics[scale=0.4]{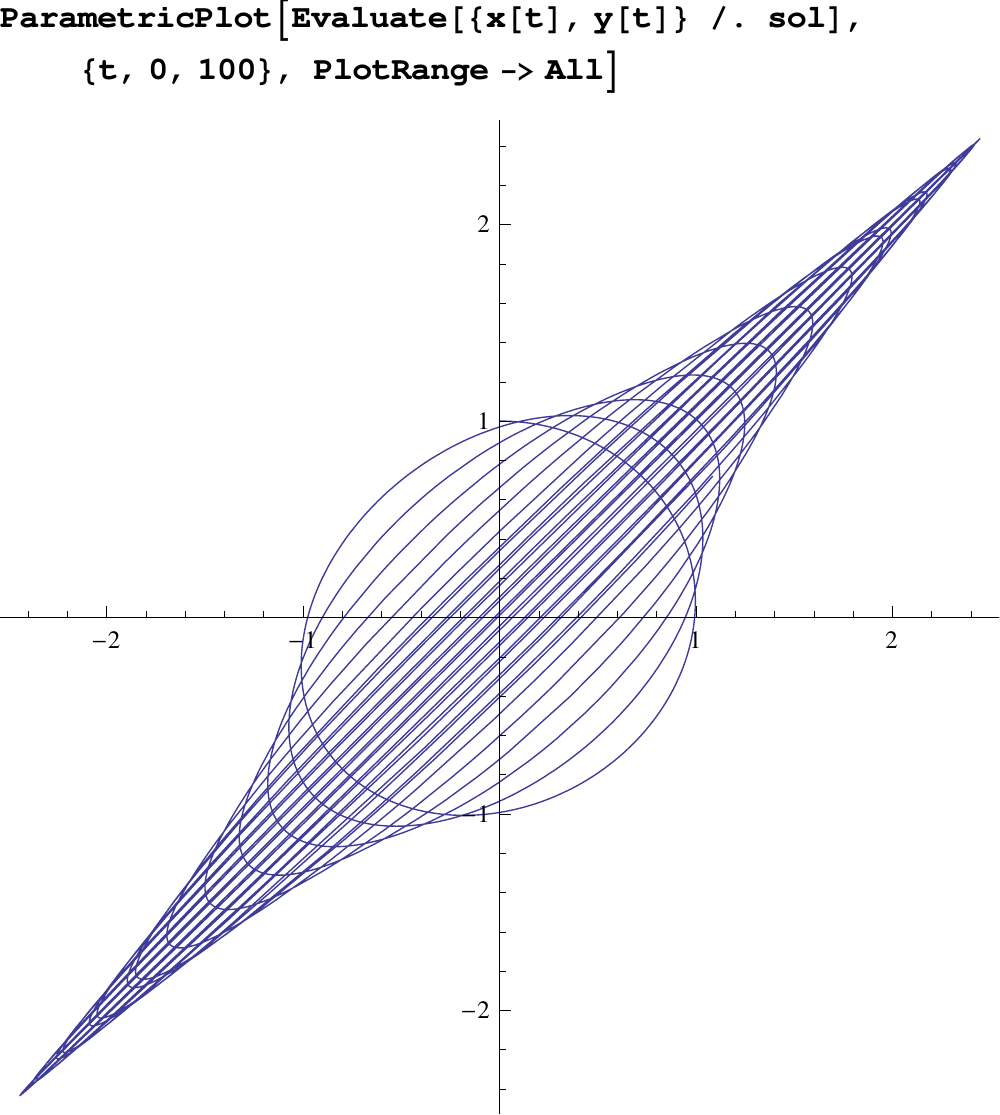}}
\put(90,61){\includegraphics[scale=0.4]{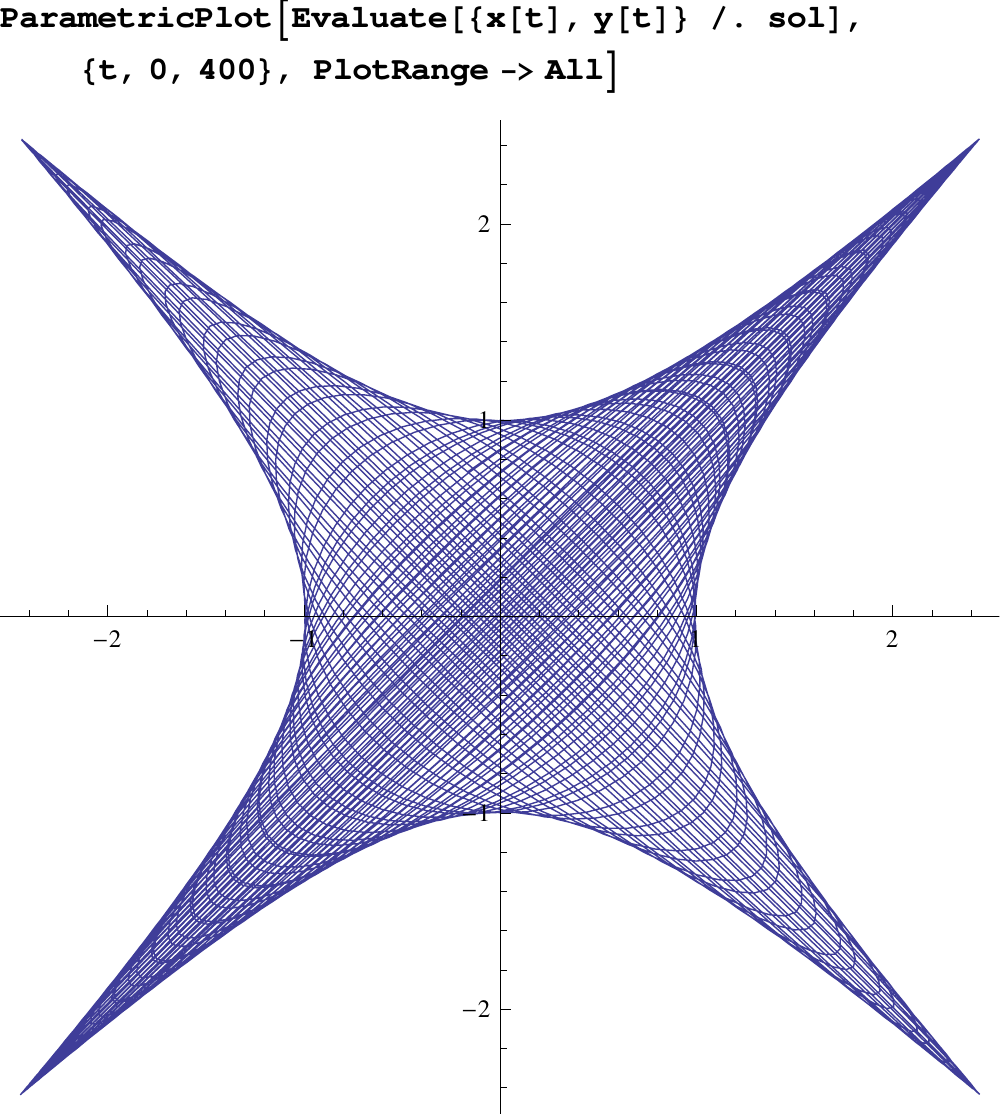}}
\put(150,61){\includegraphics[scale=0.4]{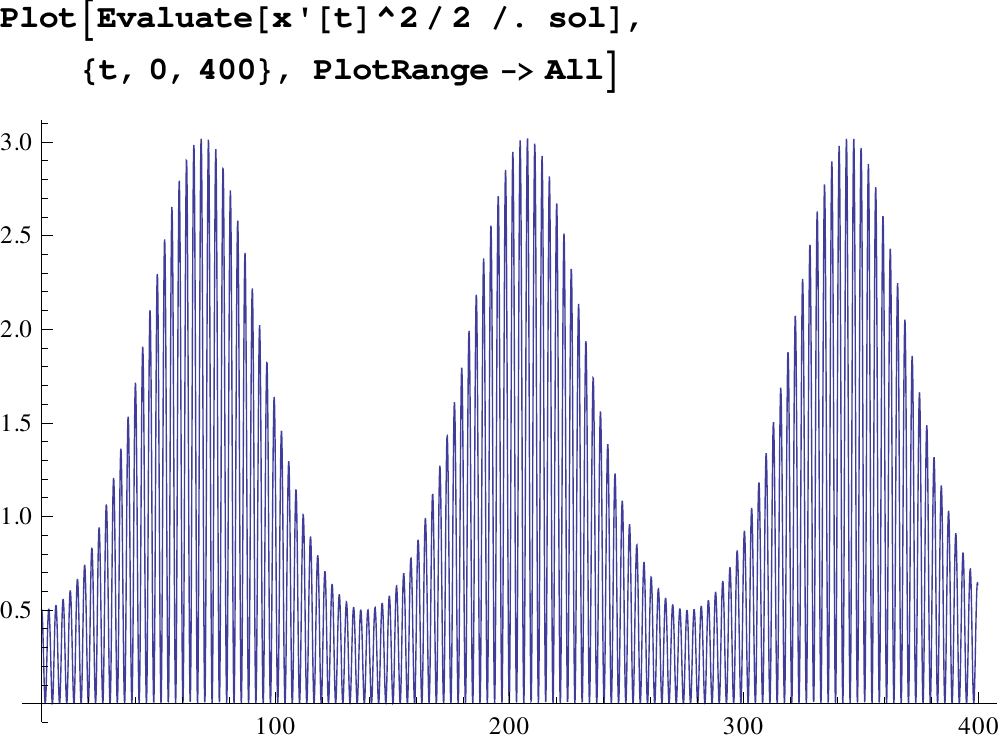}}
\put(25,0){\includegraphics[scale=0.4]{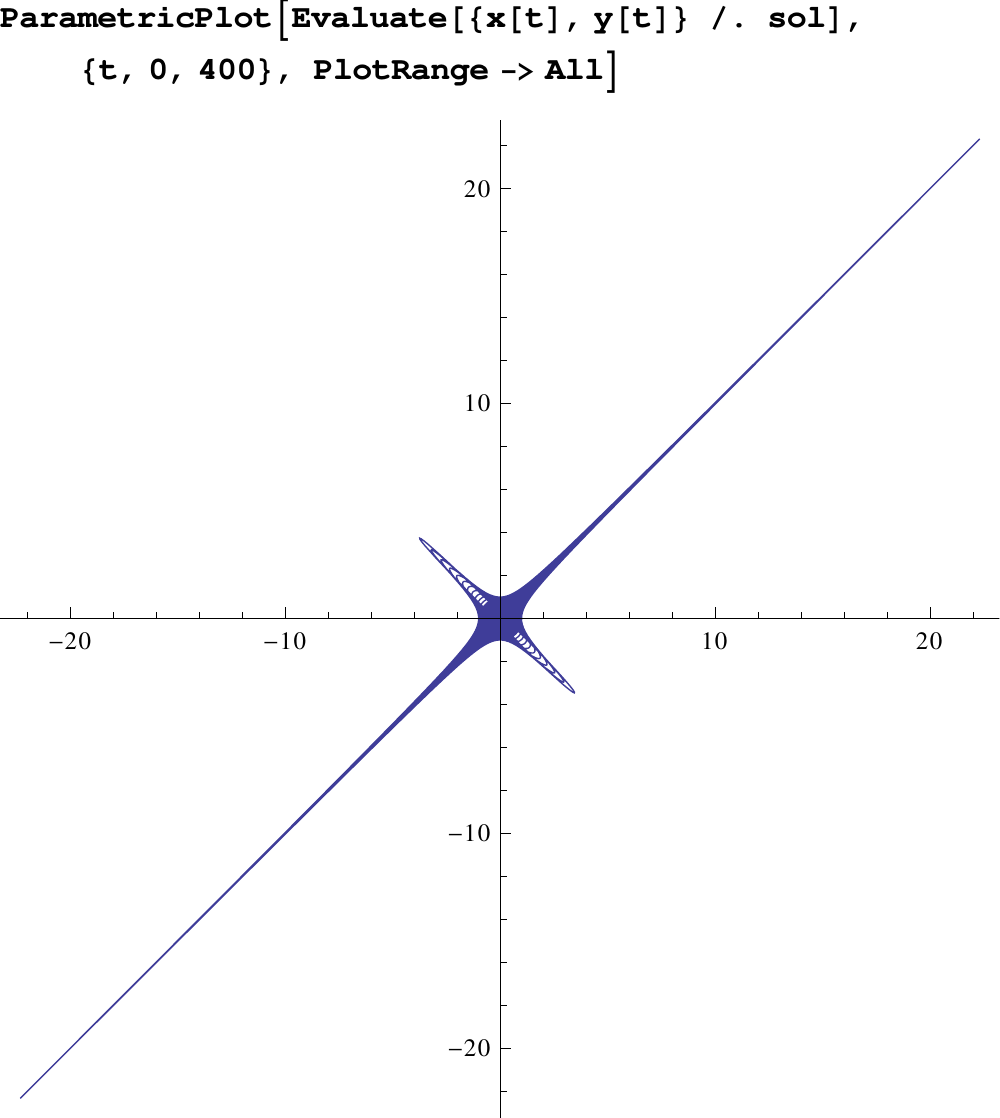}}
\put(90,0){\includegraphics[scale=0.4]{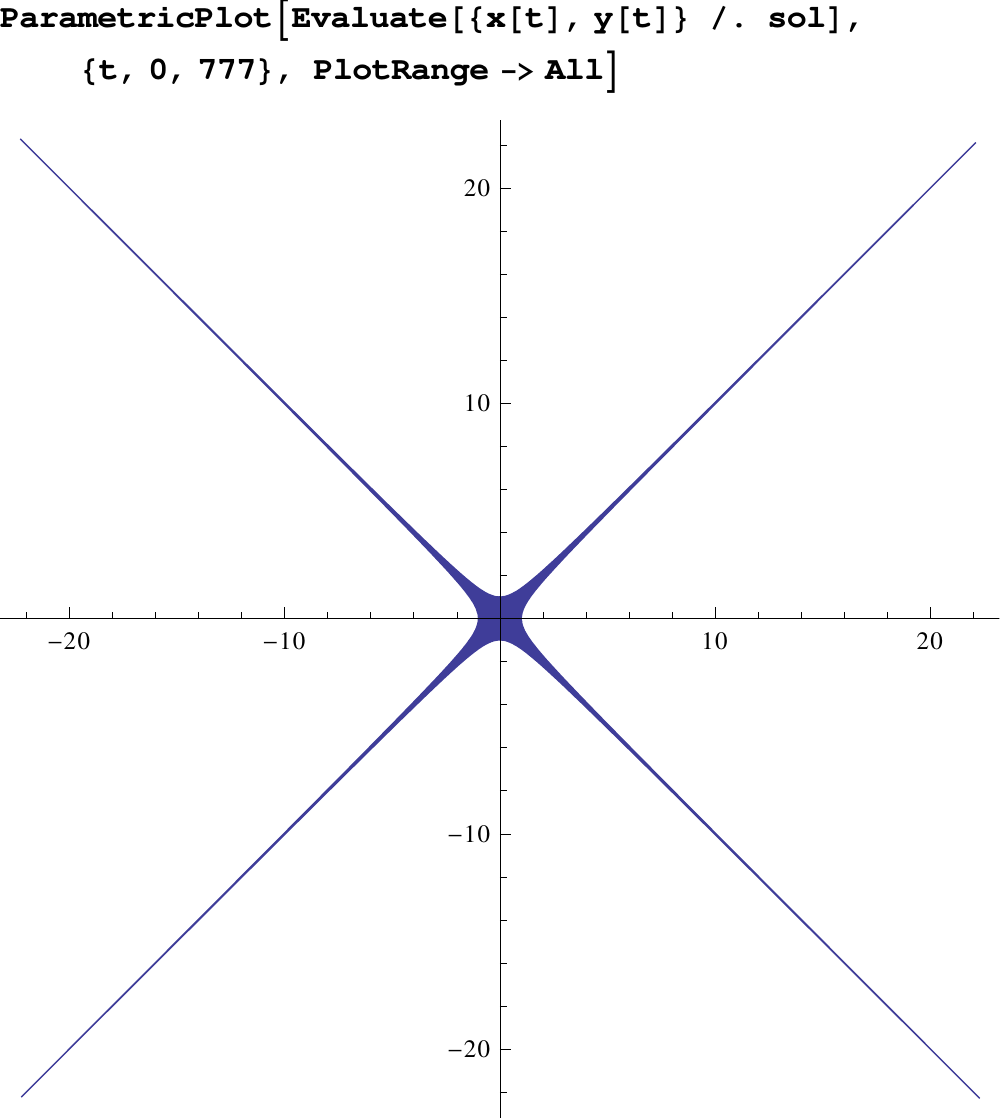}}
\put(150,0){\includegraphics[scale=0.4]{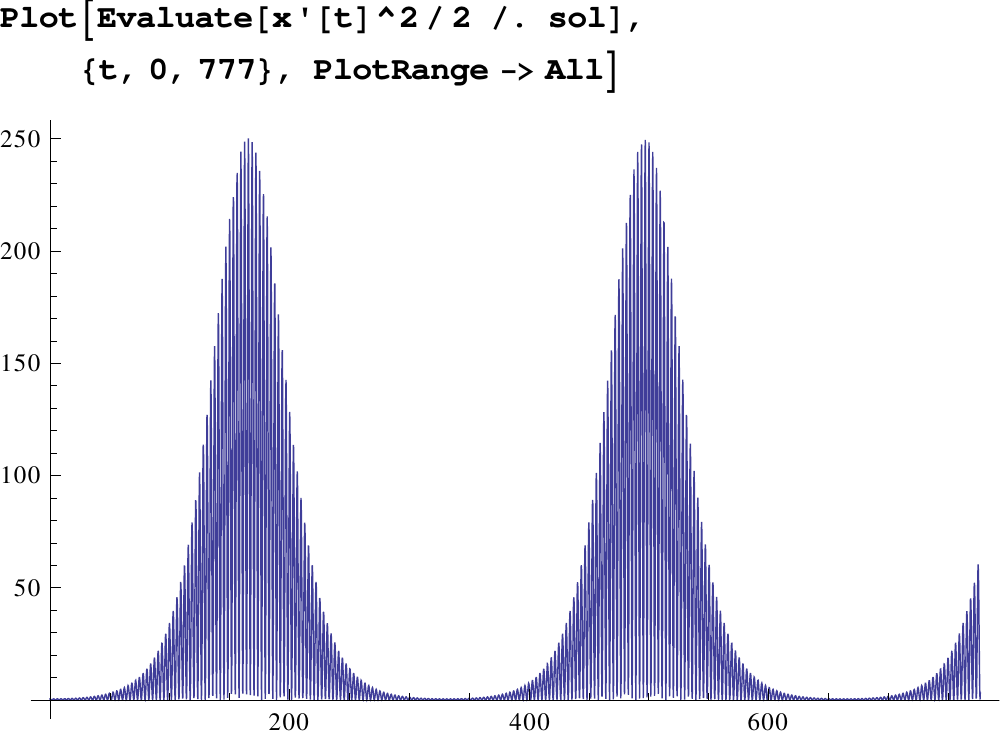}}

\put(73,82){$x$}
\put(46,110){$y$}

\put(139,82){$x$}
\put(112,110){$y$}

\put(139,21){$x$}
\put(111,48){$y$}

\put(73,21){$x$}
\put(46,48){$y$}

\put(145,92){$\frac{{\dot x}^2}{2}$}
\put(199,57){$t$}

\put(145,30){$\frac{{\dot x}^2}{2}$}
\put(201,-2){$t$}

\end{picture}

\caption{\footnotesize Up left and middle: The $(x,y)$ plot of the solution to
eqs.\,(\ref{5.7})(\ref{5.8}) for constants
$\mu=1.01$ and $\nu=1$. Up right: The kinetic energy ${\dot x}^2/2$ as
function of time. Low left and middle: The $(x,y)$ plot of the solution
to eqs.\,(\ref{5.7})(\ref{5.8}) for constants
$\mu=1.0001$ and $\nu=1$. Low right: The kinetic energy ${\dot x}^2/2$ as
function of time.}
\end{figure} 

\nnn The solution for $\mu=1.01,~\nu=1$, and the same initial condition as in
Figs.\,3,4 is shown in Fig.\,6.  We see that the trajectory in the $(x,y)$ space
does not escape into infinity. Instead, a second arm is formed, and the
trajectory remains confined within a star-like envelope. The kinetic
energy of one component, ${\dot x}^2/2$, now does not grow to infinity.
Instead, the amplitude of the kinetic energy oscillations is modulated by
a slowly oscillating envelope.

In Fig.\,6  we also repeat the calculation for $\mu=1.0001$ and $\nu=1$. This is now
very close to $\mu=1$, $\nu=1$, the case of
eqs.\,(\ref{5.5}),(\ref{5.6}). Now the trajectory in the $(x,y)$ space
goes much farther from the origin than in the case $\mu=1.01$.
  But again it does not
go to infinity; instead, after some time, the trajectory start to move within a
second arm. The envelope of the kinetic energy oscillations now consists of
separated peaks. If $\mu$ approaches $1$, the height of the peaks
becomes higher and higher, and their separation increases. In the limit
$\mu \rightarrow 1$, the height of the first peak is infinite, and there is no
second or other peaks (because their positions recedes to infinity).

We have found a very interesting result that $\mu=1$, $\nu=1$ is a singular
case, in which the system is unstable. But if $\mu$ slightly differs from
$\nu$, then the system is stable; its trajectory remains confined within
a finite region of the $(x,y)$ space, and the kinetic energy remains finite
as well.

We will now show that the stable system (\ref{5.7}),(\ref{5.8}) is a special
case of a more general system, described by the action
\be
I = \frac{1}{2}\int {\dd \tau \,g_{\mu \nu } \,\dot X^\mu  \dot X^\nu  } ~,~~~~~~
\mu,\nu = 0,1,2,...,D-1 .
\lbl{5.9}
\ee
This is the Stueckelberg action\,\ci{Stueckelberg} for a point particle in a $D$-dimensional
curved space with a metric $g_{\mu \nu}$\,\ci{PavsicBook} (see also\,\ci{HorwitzCurved}).
Introducing
\be
\gamma _{ab}  = g_{ab}  - \frac{{g_{0a} g_{0b} }}{{g_{00} }} ,
\lbl{5.10}
\ee
we can split\footnote{An analogous splitting is used in Kaluza-Klein theories.}
the quadratic form according to
\be
g_{\mu \nu } \dot X^\mu  \dot X^\nu   
= \gamma _{ab} \dot X^a \dot X^b  + \frac{{\dot X_0^2 }}{{g_{00} }} .
\lbl{5.11}
\ee
Here $a,b=1,2,...,D-1$. In general, the signature is $(r,s)$, with
$r+s=D-1$. We wil take $r=s$. Using (\ref{5.10}),(\ref{5.11}),  
the action (\ref{5.9}) then becomes
\be
I = \frac{1}{2} \int \dd \tau \,\left ( \gamma _{ab} \dot X^a \dot X^b  
    + \frac{{\dot X}_0^2 }{g_{00}} \right) .
\lbl{5.12}
\ee

If $g_{\mu \nu}=0$, then ${\dot X}_0$ is a constant of
motion\footnote{Analogously, in Kaluza-Klein theories, ${\dot X}_5$ is a
constant of motion, if the 5$D$ metric does not depend on the fifth dimension.}.
We will assume that this is the case.

Assuming that $\gam_{ab,c}\equiv (\p/\p X^c) \gam_{ab} =0$ , the equations
of motion derived from (\ref{5.12}) are
\be
\ddot X^a  + \frac{1}{2}\frac{C}{{g_{00}^2 }}\,g_{00,b} \,\gamma ^{ab}  = 0 ,
\lbl{5.14}
\ee
where $C={\dot X}_0^2$ is a constant.

Introducing
\be
V =  - \frac{1}{2}\frac{{\dot X_0^2 }}{{g_{00} }} 
=  - \,\frac{1}{2}\frac{C}{{g_{00} }} ,
\lbl{5.15}
\ee
we have
\be
\ddot X^a  + \,\,V_{,b} \,\gamma ^{ab}  = 0
\lbl{5.16}
\ee
The latter equations correspond to eqs.\,(\ref{5.7}),(\ref{5.8}), if
$\gam^{ab}={\rm diag}(\gam^{11},\gam^{22},...,\gam^{D-1,D-1})$, i.e.,
a diagonal metric. Then eqs.\,(\ref{5.16}) become
\bear
&&\ddot X^1  + \,\,V_{,1} \,\gamma ^{11}  = 0, \nonumber\\
&&\ddot X^2  + \,\,V_{,2} \,\gamma ^{22}  = 0, \nonumber\\
&& ~~ \vdots \lbl{5.18}
\ear
If we consider the case $D-1=2$, identify $X^1\equiv x$, $X^2 \equiv y$,
$\gam^{11}\equiv \mu$, $\gam^{22} \equiv \nu$, and take
$V=\frac{\omega^2}{2} (x^2-y^2) + \frac{\lambda}{4} (x^2-y^2)^2$,
then the system (\ref{5.18}) becomes the system (\ref{5.7}),(\ref{5.8}).

The $g_{\mu \nu} (x)$ in the action (\ref{5.9}) need not be a fixed
background metric field. It can be a dynamical field. Then one has to
include into the action (\ref{5.9}) a kinetic term for $g_{\mu \nu}$,
e.g., the Einstein-Hilbert term with the curvature scalar. The dynamics
of such a system is involved, and, in view of the results given in Fig.\,6,
one cannot a priori claim that the system is unstable.

\subsection{Collision of the oscillator with a free particle}

Let us assume that in the surroundings of the oscillator, $O$, described
by the Lagrangian (\ref{5.1}), there is free particle, $P$. Depending on the
initial conditions, it may happen that the oscillator hits the particle.
Such combined system of $O$ and $P$ can be modeled by the Lagrangian
\be
L = \frac{1}{2}(\dot x^2  - \dot y^2 ) - \frac{1}{2}(x^2  - y^2 ) 
- \frac{\lambda }{2}(x^2  - y^2 )^2  + \frac{1}{2}(\dot u^2  
+ \dot v^2 ) - \frac{{\alpha /5}}{{[(u - x)^2  + (v - y)^2  + a]^5 }} ,
\lbl{c1}
\ee
where $u,v$ are particle's coordinates. The equations of motions are
\be
\begin{array}{l}
 \ddot x + x + \lambda x(x^2  - y^2 ) + \frac{{\alpha (u - x)}}{{[(u - x)^2  
+ (v - y)^2  + a]^{5/2} }} = 0 \\ 
 \ddot y + y + \lambda y(x^2  - y^2 ) + \frac{{\alpha (v - y)}}{{[(u - x)^2  
+ (v - y)^2  + a]^{5/2} }} = 0 \\ 
 \ddot u - \frac{{\alpha (u - x)}}{{(u - x)^2  + (v - y)^2  + a}} = 0 \\ 
 \ddot v - \frac{{\alpha (v - y)}}{{[(u - x)^2  + (v - y)^2  + a]^{5/2} }} 
= 0 \\ 
 \end{array}
\lbl{c2}
\ee

The numerical solution for the constants $\Lambda =0.1$, $\alpha=1$, and the
initial conditions $\dot x(0) = 1\,,\,\,\,\dot y(0) = 0\,,\,\,\,\,$
$\dot u(0) = 0\,,\,\,\,\,\dot v(0) = 0$, $x(0) = 0\,,\,\,\,y(0) = 1\,,\,\,\,\,
u(0) = 12\,,\,\,\,v(0) = 11.5$ is shown in Fig.\,7. We see that the solution
properly reproduces the fact that the particle $P$ is initially at rest, and
after the interaction with the oscillator $O$, its moves with a constant
velocity.

\setlength{\unitlength}{0.8mm}

\begin{figure}[h!]
\hs{3mm}
\begin{picture}(120,90)(25,0)

\put(36,45){\includegraphics[scale=0.42]{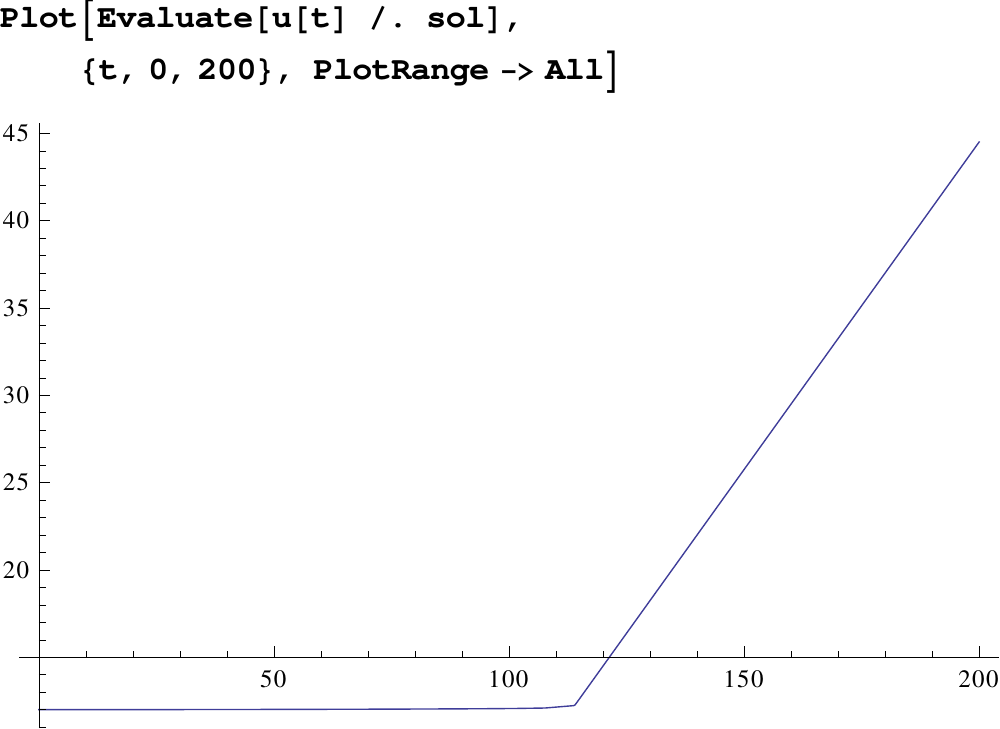}}
\put(122,45){\includegraphics[scale=0.42]{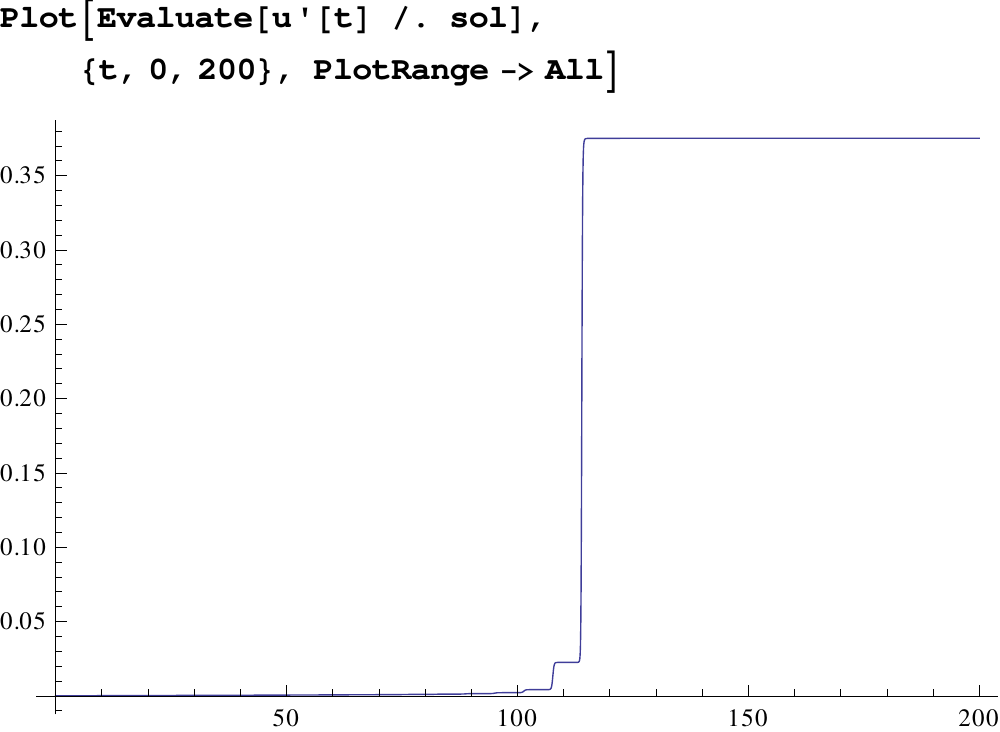}}
\put(25,0){\includegraphics[scale=0.35]{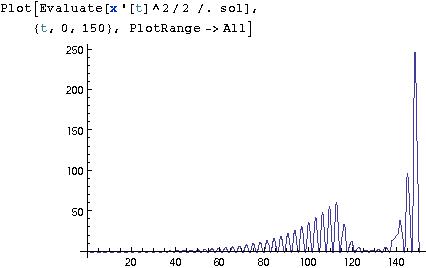}}
\put(120,0){\includegraphics[scale=0.42]{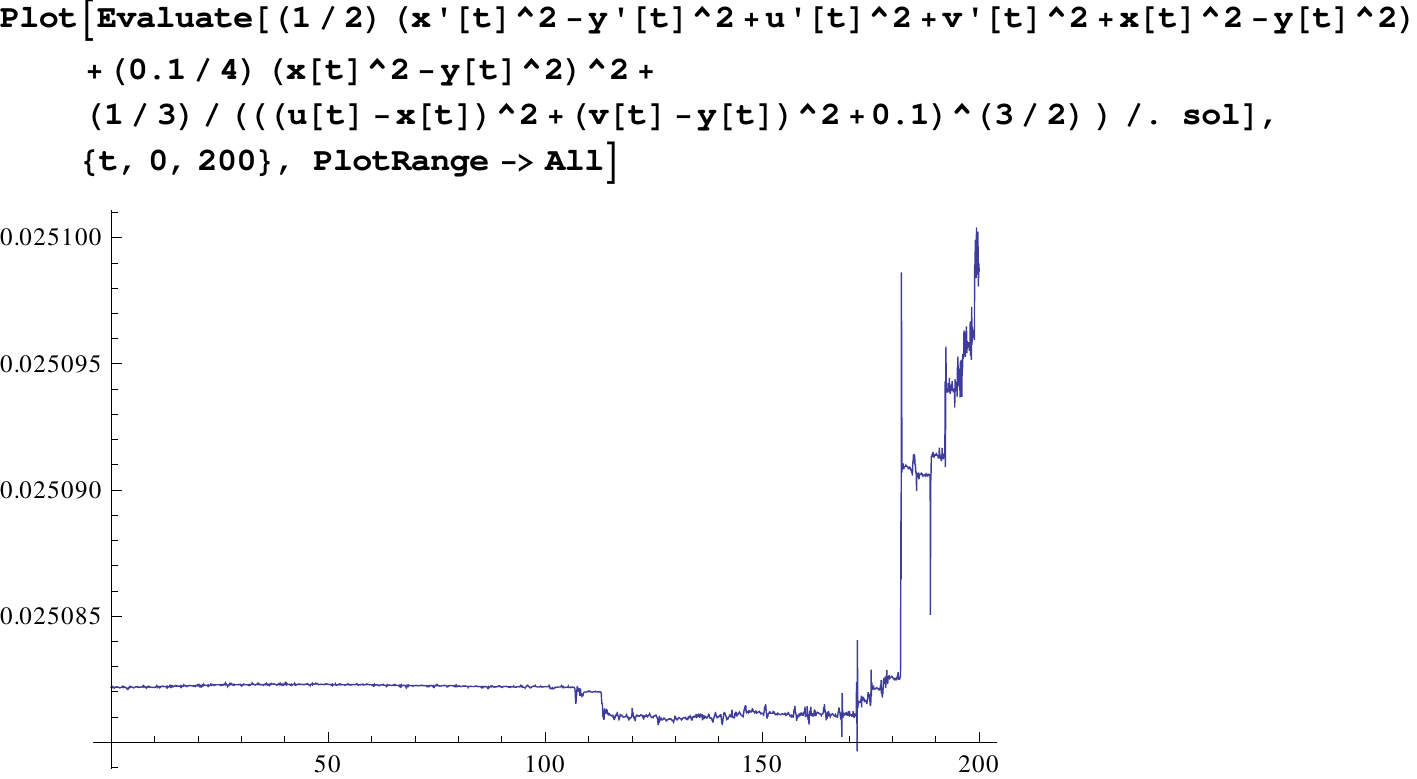}}

\put(89,44){$t$}
\put(40,75){$x$}

\put(91,-1){$t$}
\put(40,30){$\frac{{\dot x}^2}{2}$}

\put(176,44){$t$}
\put(127,75){${\dot x}$}

\put(176,-1){$t$}
\put(126,27){$E_{\rm tot}$}

\end{picture}

\caption{\footnotesize Collision of the oscillator with a particle.
Up: Particle's position and velocity as functions of time.
Low left: Oscillator's kinetic energy ${\dot x}^2/2$ as function of time.
Low left: The total energy of the the oscillator and the particle
as function of time.}
\end{figure}
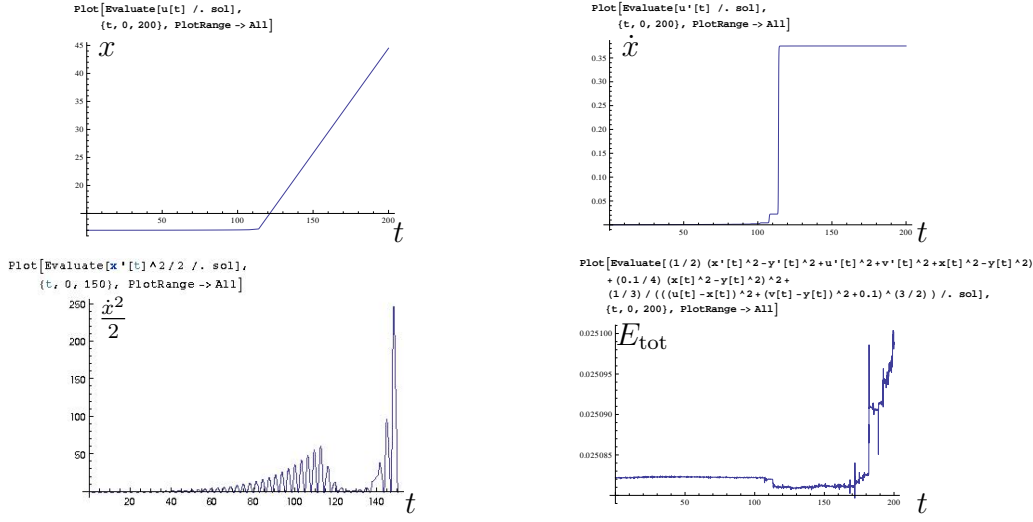 
\nnn The positive component of the oscillator's kinetic energy starts to increase,
but at time around $t=112$ it drops down to zero. After a while, the positive
kinetic energy ``recovers'' and start to increase again. A further collision
with some other particle would again drop down ${\dot x}^2/2$. Analogous
holds for ${\dot y}^2/2$. We see that, according to this numerical
solution, the surrounding particles stabilize the oscillator and prevent
it to escape into the infinity. We expect that many such oscillators, immersed
into a bath of particles would increase their average ${\dot u}^2/2$ and 
${\dot v}^2/2$, i.e., the temperature of the bath. Further investigation
of this interesting topics is beyond the scope of the present paper.

\subsection{Quantum oscillator}

The quantum oscillator is described by the wave function
$\psi(t,x,y)$, satisfying the Schr\"odinger equation
\be
\,i\frac{{\partial \psi }}{{\partial t}} = H\psi ,
\lbl{5.19}
\ee
with
\be
H = \frac{1}{2}\left( { - \,\frac{\partial }{{\partial x^2 }}
 + \,\frac{\partial }{{\partial y^2 }}} \right) + V(x,y) .
\lbl{5.20}
\ee

Let us expand the wave function in terms of the basis functions of the
2$D$ harmonic pseudo Euclidean oscillator, which are the same as those
of the Euclidean oscillator:
\be
\,\psi  = \sum\limits_{m,n=0}^\infty {c_{mn} (t)\,\psi _{mn} } .
\lbl{5.21}
\ee
Here $\psi_{mn}= \frac{1}{\pi^{1/4} \sqrt{2^n n!}}\frac{1}{\pi^{1/4} \sqrt{2^m m!}}
H_n (x) H_m (y) {\rm e}^{-(x^2 + y^2)/2}$
are orthonormal eigenfunctions (with positive or negative
energies), of the Hamiltonian
\be
H_0 = \frac{1}{2}\left( { - \,\frac{\partial }{{\partial x^2 }} 
+ \,\frac{\partial }{{\partial y^2 }}} \right) 
+ \frac{1}{2} (x^2-y^2) .
\lbl{5.22}
\ee
Using the matrix elements
\be
\,H_{mn;rs}  = \int {\rm{d}} x\,{\rm{d}}y\,\psi _{mn}^* H\psi _{rs} ,
\lbl{5.23}
\ee
the Schr\"odinger equation (\ref{5.19}) can be rewritten as
\be
i\dot c_{mn} \, = \,\sum\limits_{rs} {H_{mn;rs\,} c_{rs} } .
\lbl{5.24}
\ee

\setlength{\unitlength}{.8mm}

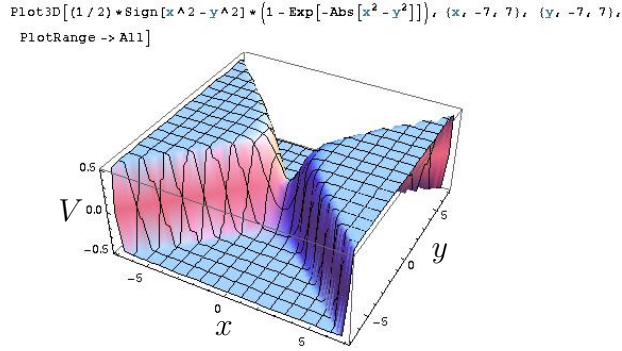
\begin{figure}[h!]
\hs{3mm}
\begin{picture}(120,50)(25,0)
\put(50,0){\includegraphics[scale=0.4]{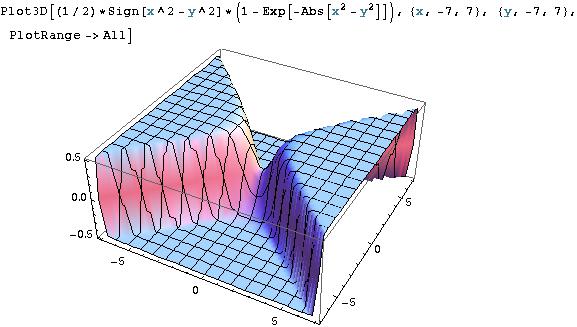}}

\put(84,2){$x$}
\put(120,15){$y$}
\put(58,21){$V$}

\end{picture}

\caption{\footnotesize The potential used in the calculations of the quantum
pseudoEuclidean harmonic oscillator described by
eqs.\,(\ref{5.19}),(\ref{5.20}).}
\end{figure} 

We will investigate the case of the potential (see Fig.\,8)
\be
V(x,y) = \frac{1}{2}\varepsilon \left( {\,1 
- e^{ - \varepsilon (x^2  - y^2 )} } \right) ,
\lbl{5.25}
\ee
\nnn where $\epsilon = {\rm sign}\, (x^2-y^2)$, i.e., $\epsilon =1$, if $x^2-y^2 >0$,
$\epsilon =-1$, if $x^2-y^2 < 0$, and $\epsilon =0$, if $x^2-y^2 =0$.
Such a potential is a 2D model of a potential that could eventually occur
in a three or higher dimensional world, where potentials running into
infinity are unrealistic.

\begin{figure}[h!]
\hs{3mm}
\begin
{picture}(120,100)(25,0)
\put(15,55){\includegraphics[scale=0.35]{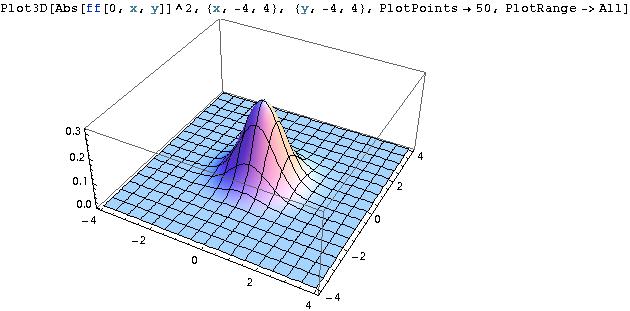}}
\put(120,55){\includegraphics[scale=0.35]{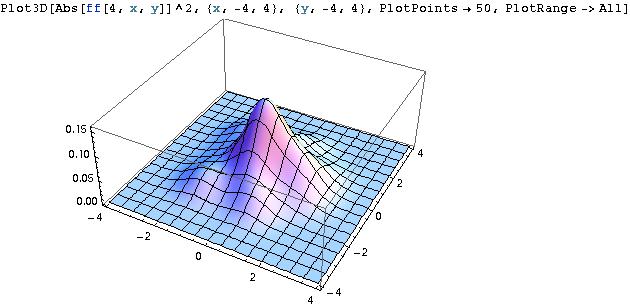}}
\put(120,0){\includegraphics[scale=0.35]{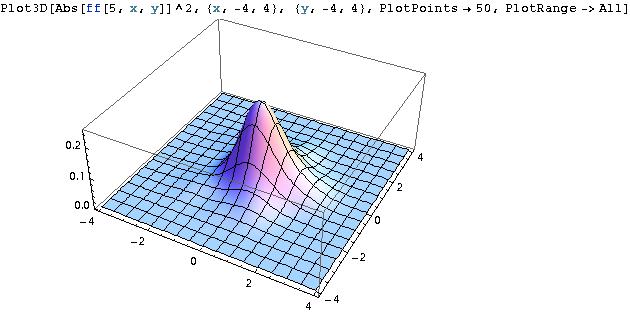}}

\put(88,90){$t=0$}
\put(191,90){$t=4$}
\put(191,35){$t=5$}

\put(44,58){$x$}
\put(75,65){$y$}
\put(16,75){$|\psi|^2$}

\end{picture}

\caption{\footnotesize The plot of $|\psi(t,x,y)|^2$, calculated for the initial conditions
$c_{00}(0)=1$, $c_{01}(0)=C_{10}(0)=c_{11}(0) = ... = 0$, at different values of
the time $t$.}
\end{figure}
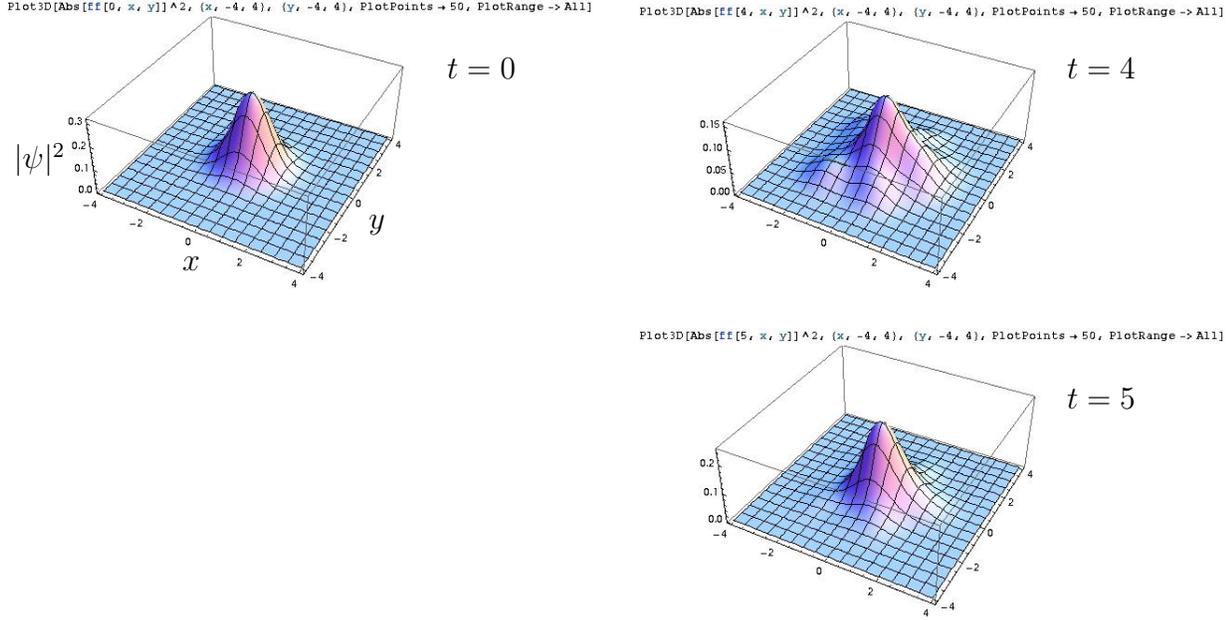 

Instead of the system (\ref{5.24}) of infinite number of first order
differential equations, let us consider a finite system with
$m,n=0,1,2,...,N$. Then we can numerically solve the system and
calculate the coefficients $C_{mn}(t)$ for given initial conditions.
So we obtain the wave functions
\be
\,\psi(t,x,y)  = \sum\limits_{m,n=0}^N {c_{mn} (t)\,\psi _{mn} } ,
\lbl{5.25a}
\ee
and its absolute square, $|\psi(t,x,y)|^2$.

In Fig.\,9  we show the result for $N=4$, and the initial condition
$c_{00}=1$, with the remaining coefficients being equal to zero.
We see that with increasing time $t$, the vacuum $\psi(0)=\psi_{00}=
    $ gradually decays, and, after a while (at $t\approx 5$) occurs
again. The wave function thus oscillates between the vacuum and
a decayed vacuum.

In Fig.\,10  we show the calculation of $|\psi(t,x,y)|^2$ for the initial
conditions $c_{01}(0)=1/\sqrt{2}$, $c_{10}(0)=1/\sqrt{2}$, 
$c_{00}(0)=0$, $c_{11}(0)=0$, $c_{12} (0)=0,..., c_{44}(0)=0$.
Initially, we have two peaks, one for the positive energy excitation,
$c_{10}(0)$, and the other one for the negative energy excitations,
$c_{01}(0)$. The system then oscillates as shown in Fig.\,10.

\setlength{\unitlength}{.8mm}

\begin{figure}
\hs{3mm}
\begin{picture}(120,240)(25,0)
\put(25,200){\includegraphics[scale=0.35]{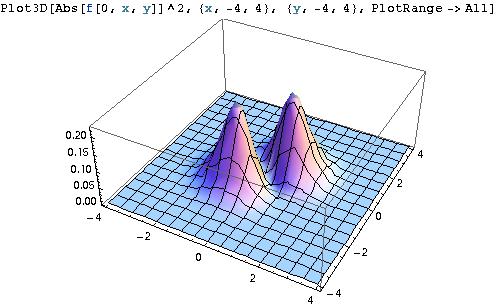}}
\put(120,200){\includegraphics[scale=0.35]{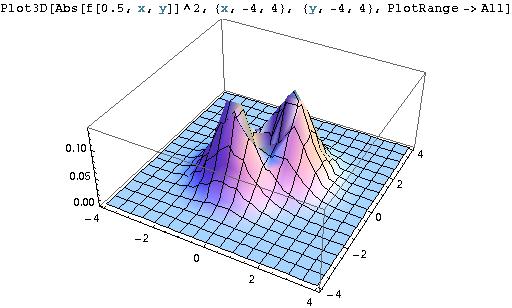}}
\put(25,150){\includegraphics[scale=0.35]{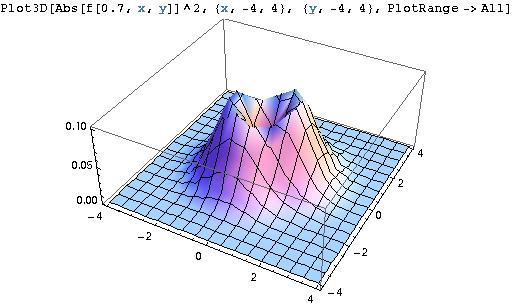}}
\put(120,150){\includegraphics[scale=0.35]{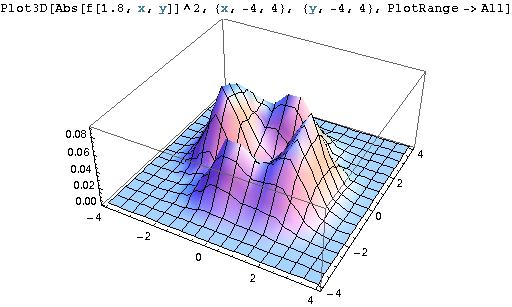}}

\put(25,100){\includegraphics[scale=0.35]{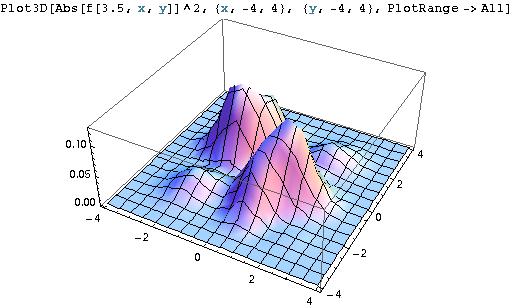}}
\put(120,100){\includegraphics[scale=0.35]{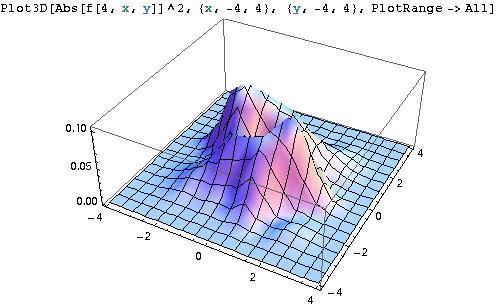}}
\put(25,50){\includegraphics[scale=0.35]{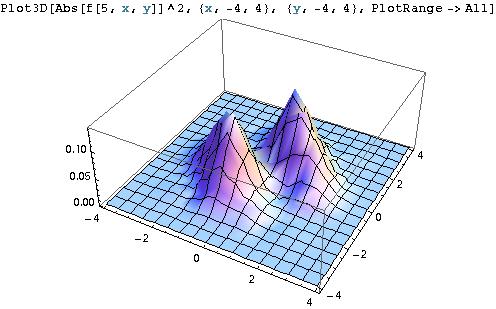}}
\put(120,50){\includegraphics[scale=0.35]{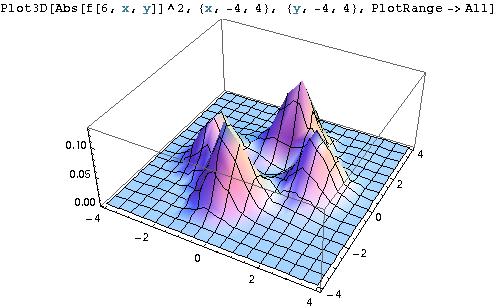}}
\put(25,0){\includegraphics[scale=0.35]{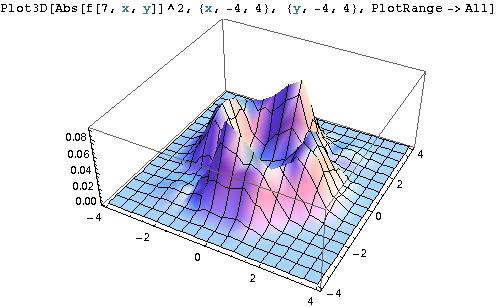}}
\put(120,0){\includegraphics[scale=0.35]{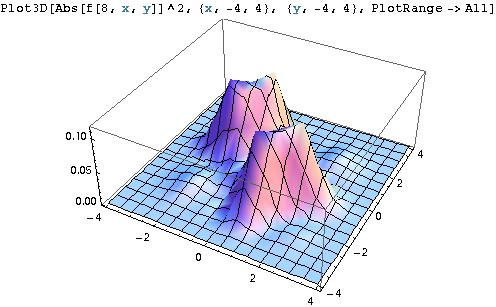}}

\put(52,204){$x$}
\put(85,211){$y$}
\put(26,220){$|\psi|^2$}

\end{picture}

\caption{\footnotesize The plot of $|\psi(t,x,y)|^2$, calculated for the initial conditions
 $c_{01}(0)=C_{10}(0)=1/\sqrt{2}$, $c_{00}(0)=c_{11}(0)=c_{12}(0) = ... = 0$,
 at different values of
the time $t$. }
\end{figure} 

\setlength{\unitlength}{.8mm}

\begin{figure}
\hs{3mm}
\begin{picture}(120,100)(25,0)

\put(25,45){\includegraphics[scale=0.35]{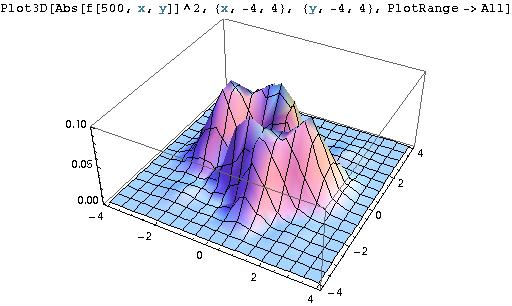}}
\put(120,45){\includegraphics[scale=0.35]{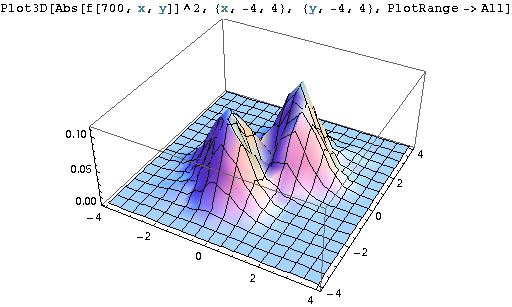}}
\put(25,0){\includegraphics[scale=0.35]{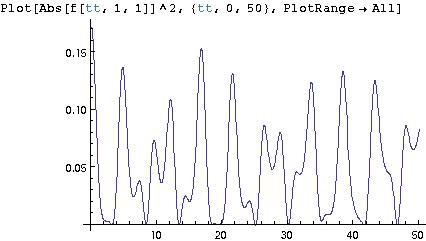}}
\put(120,0){\includegraphics[scale=0.35]{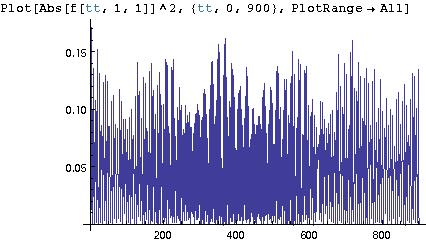}}

\put(52,49){$x$}
\put(85,55){$y$}
\put(26,65){$|\psi|^2$}

\put(148,49){$x$}
\put(180,55){$y$}
\put(121,65){$|\psi|^2$}

\put(27,29){$|\psi|^2$}
\put(92,-1){$t$}

\put(122,29){$|\psi|^2$}
\put(186,-1){$t$}

\end{picture}

\caption{\footnotesize Up: The plot of $|\psi(t,x,y)|^2$, calculated for the initial conditions
 $c_{01}(0)=C_{10}(0)=1/\sqrt{2}$, $c_{00}(0)=c_{11}(0)=c_{12}(0) = ... = 0$,
 at two very large values the time, $t=500$ and $t=700$.
 Low: The plot of $|\psi(t,x=0,y=50)|^2$, as a function of $t$, calculated for the
 same initial conditions as above.}
\end{figure} 

Such a wave function (\ref{5.26}), of course, is not a solution of
the Schr\"odinger equation (\ref{5.19}) for $H$ given in eqs.\,(\ref{5.20}),
(\ref{5.25}). Since $c_{mn} (t)$, $m,n=0,1,2,3,4$, are solution of the
truncated system (\ref{5.24}), also $\psi(t,x,y)$ of eq.\,(\ref{5.25a})
is a solution of a truncated system, presumably of the Schr\"odinge equation
on the lattice. A realistic lattice is a discrete set of closely
separated points corresponding to the system, not of the $5\times 5$ equations
(\ref{5.24}), but of the $N \times N$ equations, with $N$ very large.
The $N$ is related to the maximum absolute energy of the oscillator
excitations that, in turn, is related to the minimum distance (cutoff).
Regardles of how large is $N$, and how small is the cutoff distance, the
examples of Figs.\,9--11  indicate that there are oscillations, and that the
system is thus stable. Only in the limit of infinitely small cutoff
distance, implying infinite quantum numbers $m,n$ of poisitive and negative
energy excitations, would the system be unstable. Then the peak in Fig.\,9
would spread to infinity. However, according to common belief, the Planck
length is the cutoff distance under which it is no longer possible to
probe spacetime distances. Our spacetime is then like a huge lattice.
Then also the corresponding Clifford space, $C$, is like a lattice. The
space $M_{1,1}$ is a subspace of $C$. Instead of the infinite system
of differential equations (\ref{5.24}), we then have a finite, though very
big, system of differential equations.

\subsection{Interacting quantum fields}

\subsubsection{Scalar fields}

As an example let us consider the scalar fields described by the action
\be
   I = \frac{1}{2}\int {\dd^4 x [g^{\mu \nu } \partial _\mu  \varphi ^a 
   \partial _\nu  \varphi ^b \gam_{ab}  + V(\varphi )]} , 
\lbl{5.26}
\ee
which is similar to the action (\ref{4.1}), except that now $V(\varphi)$
is a more general potential with an interaction term, e.g.,
$V(\varphi)= m^2 \varphi^a \varphi^b \gam_{ab}
 + \varphi^a \varphi^b \varphi^c \varphi^d \lambda_{abcd}$.

Upon quantization, the system is described by a state vector,
\be
|\Psi \rangle  = \sum\limits_{} {|P\rangle \,\langle P|\Psi \rangle } ,
\lbl{5.27}
\ee
that can be expanded in terms of the Fock space basis vectors
$|P \rangle \equiv |p_1 p_2 ... p_n \rangle$ that are eigenvectors of the
free field Hamiltonian $H_0$ (without the interaction quartic term
$\varphi^a \varphi^b \varphi^c \varphi^d \lambda_{abcd}$). Also the vacuum,
$|0 \rangle$, is an eigenstate $H_0$.
If the metric $\gam_{ab}$ has signature
$(r,s)$, the states $|P \rangle$ can have positive or negative energies.

The time evolution of the state vectors is governed by the Hamilton operator
$H$, corresponding to the field action (\ref{5.26}):
\be
  |\Psi (t)\rangle  = e^{ - iH(t - t_0 )} |\Psi (t_0 )\rangle .
\lbl{5.28}
\ee
From (\ref{5.28}) we have
\be
\langle P|\Psi (t)\rangle  = \sum\limits_{P'} {\langle P|e^{ - iH(t - t_0 )}
 |P'\rangle \,\langle P'\,|\Psi (t_0 )\rangle } .
\lbl{5.29}
\ee
If the initial state is the vacuum, $|\psi (t_0) \rangle =|0 \rangle$,
an eigenstate of the unperturbed Hamiltonian, $H_0$,
then we have
\be
\langle P|\Psi (t)\rangle  = \langle P|e^{ - iH(t - t_0 )} |0\rangle ,
\lbl{5.30}
\ee
Such transition is possible, because a state $\langle P|$ contains particles
with positive and negative energies. Therefore, the total energy in such
transition is conserved and remains equal to the energy of the initial state
$|0 \rangle$, which is zero.

The vacuum thus decays into a superposition of many particle states:
\be
 |\Psi (t)\rangle  = \sum\limits_{n = 0}^\infty  
 {|p_1 p_2 ...p_n \rangle \,\langle p_1 p_2 ...p_n |\Psi (t)\rangle } .
\lbl{5.31}
\ee
Here $\langle p_1 p_2 ...p_n |\Psi (t)\rangle$ is the amplitude that we will
measure the multiparticle state $|p_1 p_2 ...p_n \rangle$. The probabilities
that the vacuum decays into any of the states $|p_1 \rangle$,
$|p_1 p_2 \rangle$, $|p_1 p_2 ...p_n \rangle$, ...,
are not drastically different from each other, and they sum to $1$:
\be
\sum\limits_{p_1 } {|\langle p_1 |\Psi \rangle |^2  + } 
\sum\limits_{p_1 ,p_2 } {|\langle p_1 ,p_2 |\Psi \rangle |^2  + } 
\sum\limits_{p_1 ,p_2 ,...,p_n } {|\langle p_1 ,p_2 ,...,p_n |\Psi \rangle |^2
  + } \,...\,\, = \,1 .
\lbl{5.32}
\ee
We see that the probability of vacuum decay into 2,4,6,8, or any finite
number of particles,  is negligible in comparison to the probability of the
decay into infinite number of particles, because such configuration spaces occupy
the vast majority of the terms in the sum (\ref{5.32}).
According to such reasoning, the vacuum instantly decays into infinitely many particles.
It is then argued that because such instantaneous vacuum decay
makes no sense in physics, ultrahyperbolic spaces are
unphysical. In my opinion such conclusion is not unavoidable, for the
reasons described below and in the remaining parts of the paper.

Let us consider a generalization of the field action (\ref{5.26}).
We can rewrite it in a more compact notation:
\be
I = \frac{1}{2}\partial _\mu  \varphi ^{a(x)} \partial _\nu  \varphi ^{b(x')} 
\gamma _{\,\,\,\,\,\,\,\,a(x)b(x')}^{\mu \nu }  - U[\varphi ]
\lbl{5.33}
\ee
Here $\varphi^{a(x)} \equiv \varphi^a (x)$, where ${(x)}$ is the continuous
index, denoting components of an infinite dimensional vector. In addition,
for every $(x)$, the components are also denoted by a discrete index $a$.
Altogether, vector components are denoted by the double index $a(x)$.
Alternative notation, often used in the litarature, is
$\varphi^a (x) \equiv \varphi^{ax}$ or $\varphi^a (x) \equiv \varphi^{(ax)}$.

The action (\ref{5.33}) may be obtained from a higher dimensional action
\be
I_\phi   = \frac{1}{2}\partial _\mu  \phi ^{A(x)} \partial _\nu  
\phi ^{B(x')} \,G_{\,\,\,\,\,A(x)B(x')}^{\mu \nu } ,
\lbl{5.34}
\ee
where $A=(a,{\bar A})$, and $\phi^{A(x)} =(\phi^{a(x)},\phi^{{\bar A}(x)})$.
The higher dimensional metric is a functional of $\phi^{A(x)}$.
Performing the Kaluza-Klein split,
\be
G_{\,\,\,\,\,\,\,AB}^{\mu \nu }  = \left( \begin{array}{l}
 {\gamma^{\mu \nu }}_{ab}  + {A_a}^{\bar A} {A_b}^{\bar B} 
 {{\bar G} ^{\mu \nu }}_{\bar A\bar B} \,,\,\,\,\,\,\,\,\,\,\,\,
 {A_a}^{\bar B} {{\bar G}^{\mu \nu }}_{\bar A\bar B} \, \\ 
 \,\,\,\,{A_b}^{\bar B} {{\bar G}^{\mu \nu }}_{~\bar A\bar B} 
 \,,\,\,\,\,\,\,\,\,\,\,\,\,\,\,\,\,\,\,\,\,\,\,\,\,\,\,
 \,\,\,\,\,\,\,{{\bar G}^{\mu \nu }}_{~\bar A\bar B} \,\,\, \\ 
 \end{array} \right) ,
\lbl{5.35}
\ee
where for simplicity we have omitted the index $(x)$, we find,
\be
I_\phi   = \frac{1}{2}\partial _\mu  \phi ^{a(x)} \partial _\nu  \phi ^{b(x')}
 \gamma _{\,\,\,\,\,\,\,\,a(x)b(x')}^{\mu \nu }  
 + \frac{1}{2}\partial _\mu  \phi _{{\bar A}(x)} 
 \,\partial _\nu  \phi _{{\bar B}(x)}
  \,{\bar G}^{\mu \nu \bar A (x) \bar B (x)} .
\lbl{5.36}
\ee
Identifying $\phi^{a(x)} \equiv \varphi^{a(x)}$, and
denoting $\frac{1}{2}\partial _\mu  \phi _{\bar A (x)} 
\,\partial _\nu  \phi _{\bar B (x)}
  \,{\bar G}^{\mu \nu \bar A (x) \bar B (x)} = - U[\varphi]$,
we obtain the action (\ref{5.33}).

Thus, the field action (\ref{5.26}) is embedded in a higher dimensional action
with a metric $G_{\,\,\,\,\,\,A(x) B(x)}^{\mu \nu }$ in field space. A question
arises as to which field space metric to choose. The lesson from general relativity
tells us that the metric itself is dynamical. Let us therefore assume that this
is so in the case of field theory\ci{PavsicBook} as well. Then (\ref{5.34})
must be completed by a kinetic term, $I_G$, for the field space
metric\,\ci{PavsicBook}. The total action is then
\be
  I[\phi ,G] = I_\phi   + I_G .
\lbl{5.37}
\ee
According to such dynamical principle, not only the field $\phi$, but
also the metric $G_{\,\,\,\,\,\,A(x) B(x)}^{\mu \nu }$ changes with the
evolution of the system. This implies that also the potential $U[\varphi]$
of eq.\,(\ref{5.33}) changes with evolution, and so does the potential
$V(\varphi)$, occurring in eq.\,(\ref{5.26}).

Let us assume that the action (\ref{5.37})   describes the whole
universe\footnote{Of course, such a model universe is not realistic, because our
universe contains fermions and accompanying gauge fields as well.}.
Then $|\psi (t) \rangle$ of eq.\,(\ref{5.31}) contains everything in such
universe, including observers. There is no external observer,
${\cal O}_{\rm ext}$,
according to whom the coefficients $\langle p_1 ...p_n|\psi (t) \rangle$
in eq.\,(\ref{5.31}) could be related to the probability densities
$|\langle p_1 ...p_n|\psi (t) \rangle|^2$ of finding the system in an
$n$-particle states with momenta $p_1$, ..., $p_n$, $n=0,1,2,3,...,\infty$.
There are only inside observers, ${\cal O}$, incorporated within 
appropriate multiparticles states $|p_1 ...p_n \rangle$, $n=0,1,2,3,...,\infty$,
of the ``universal" state $|\psi (t) \rangle$. According to the Everett
interpretation of quantum mechanics, all states,  $|0 \rangle$,
$|p_1 p_2 \rangle$,..., $|p_1 ... p_n \rangle$, ..., $n=0,1,2,...,\infty$,
in the superposition (\ref{5.31}) actually occur, each in a different world.
The Everett interpretation is now getting increasing support among
cosmologists (see, e.g., Ref.\,\ci{Many worlds}).
For such an
inside observer, ${\cal O}$, there is no instantaneous vacuum decay into
infinitely many particles. For ${\cal O}$, at a given time $t$, there exists a
configuration $|P \rangle$ of $n$-particles (that includes ${\cal O}$
himself), and a certain field potential
$V(\varphi)$ (coming from ${G^{\mu \nu}}_{A(x)B(x)}$).
At some later time, $t+\Delta t$, there exists a slightly different
configuration $|P' \rangle$ and potential $V'(\varphi)$, etc.
Because ${\cal O}$ nearly continuously measures the state $|\psi (t) \rangle$ of
his universe, the evolution of the system is being ``altered", due to the notorious
``watchdog effect" or ``Zeno effect"\ci{Sudarshan} of quantum mechanics (besides being altered by
the evolution of the potential $V(\varphi)$). The peculiar behavior of a quantum
system between two measurements has also been investigated
in Refs.\,\ci{Aharonov}.

\subsubsection{The generalized Dirac field}

In Sec.\,4.2.2 we considered the generalized Dirac field described by the
Dirac-K\"ahler equation (\ref{4.13}). In the spinor basis of $Cl(1,3)$,
the field can be represented by a $4 \times 4$ matrix
\be
\psi  = \left( \begin{array}{l}
 \psi ^{11} \,\,\psi ^{12} \,\,\psi ^{13} \,\,\psi ^{14}  \\ 
 \psi ^{21} \,\,\psi ^{22} \,\,\psi ^{23} \,\,\psi ^{24}  \\ 
 \psi ^{31} \,\,\psi ^{32} \,\,\psi ^{33} \,\,\psi ^{34}  \\ 
 \psi ^{41} \,\,\psi ^{42} \,\,\psi ^{43} \,\,\psi ^{44}  \\ 
 \end{array} \right) ,
\lbl{5.38}
\ee
where each column represents a Dirac spinor\footnote{
A column may represent the Weyl and Majorana spinors as well}.
The components $\psi^{\alpha i}$ have either positive or negative
energies, according to the following scheme:
\be
{\rm{Energy}} = \left( \begin{array}{l}
  + \,\,\,\,\, + \,\,\,\,\, - \,\,\,\,\, -  \\ 
  + \,\,\,\,\, + \,\,\,\,\, - \,\,\,\,\, -  \\ 
  - \,\,\,\,\, - \,\,\,\,\, + \,\,\,\,\, +  \\ 
  - \,\,\,\,\, - \,\,\,\,\, + \,\,\,\,\, +  \\ 
 \end{array} \right) ,
\lbl{5.39}
\ee
which is a consequence of the metric (\ref{4.20}),(\ref{4.21}).

If we add an interaction term to the action (\ref{4.19}), then transitions
between positive and negative energy states become possible. On the other
hand, positive and energy states of the usual Dirac spinors do not
mix in our Universe. Even if once they did mix, the evolution of the
Universe must have led to the current situation with no mixing.
This was not so clear when Dirac proposed his theory. The occurrence of
negative energy states was puzzling at that time, and Fermi\,\ci{Fermi}
wrote:

\baselineskip 0.45cm

{\footnotesize 

"It is well known that the most serious difficulty in Dirac's
relativistic wave
equations lies in the fact that it yields besides the normal positive
states also negative ones, which have no physical significance. This would
do no harm if no transition between positive and negative state were
possible (as are, e.g., transitions between states with symmetrical and
antisymmetrical wave function). But this is unfortunately not the case:
Klein has shown by a very simple example that eleectrons impinging against
a very high potential barrier have a finite probability of going over
in a negative state."}

\baselineskip 0.55cm

Thus, Dirac's relativistic wave equation could have been put aside and
ignored. Fortunately, this did not happen. The problem was resolved by the
Dirac sea of negative enery states.

Now the situation is analogous as in 1932. Within the Clifford algebra framework,
besides the negative energy states of the first left ideal of $Cl(1,3)$,
we have also negative energy states of the second, third and forth ideal.
The existence of such states also should not be considered a priori as
problematic. A deeper investigation is necessary before making any
definite conclusion.

\subsubsection{Clifford algebra description of fermionic fields}

It is known that spinors can be represented as Fock space like
objects\,\ci{SpinorBasis,PavsicSpinorInverse,PavsicPhaseSpace}
embedded in a Clifford algebra.
An analogous procedure can be carried out for spinor fields\,\ci{PavsicBook, PavsicPhaseSpace}.
A vector $\Psi$ in an infinite dimensional space, ${\cal S}$, can be represented
as\,\ci{PavsicBook, PavsicPhaseSpace}
\be
\Psi  = \psi^{r(x)} h_{r(x)} ~,~~~~r=1,2;~~~x\in {\mathbb R}^3~{\rm or}~
       x\in {\mathbb R}^{1,3} .
\lbl{5.40}
\ee
Here $\psi^{r(x)}$ are the vector components (see Sec.\,6.4.1  ), and
$h_{r(x)}$ are basis vectors, represented as generators of the infinite
dimensional Clifford Algebra, $Cl(\infty)$, satisfying
\be
   h_{r(x)} \cdot h_{r'(x')} \equiv \frac{1}{2} (h_{r(x)} h_{r'(x')}+
   h_{r'(x')} h_{r(x)}) = \rho _{r(x)r'(x')} ,
\lbl{5.41}
\ee
where $\rho _{r(x)r'(x')}$ is the metric of ${\cal S}$. In particular,
it can be $\rho _{r(x)'(x')}=\delta_{rr'} \delta (x-x')$.
The 4-component spinor indices, $\alpha=1,2,3,4$, and the index $i=1,2,3,4$,
denoting the ideals, is implicit in $\psi^{r(x)} \equiv
\psi^{\alpha i r(x)}$, and $h_{r(x)}\equiv h_{\alpha i r(x)}$.
In a new basis, called the {\it Witt basis},
\bear
&&h_{\,\,(x)}  = \frac{1}{{\sqrt 2 }}(h_{1(x)}  + i\,h_{2(x)} ) \lbl{5.42}\\ 
 &&h_{\,*(x)} \,\, = \frac{1}{{\sqrt 2 }}(h_{1(x)}  - i\,h_{2(x)} ), \lbl{5.43}
\ear
instead of (\ref{5.41}), we have\footnote{
Instead of the indices $1,2$, denoting real and imaginary componet, we now use
the space,~, and the star, $*$.}
\bear
  &&h_{\,(x)} \cdot h_{\,*(x')}  = \rho _{\,(x)\,*(x')}  \lbl{5.44}\\ 
  &&h_{\,(x)} \cdot h_{\,(x')}  = h_{\,*(x)} \cdot h_{\,*(x')} \, = 0 , \lbl{5.45}
\ear
which are {\it fermionic anticommutation relations}. A vector $\Psi$
can be expanded as
\be
\Psi  = \psi ^{(x)} h_{\,(x)} \, + \psi ^{*(x)} h_{\,*(x)}^{} .
\lbl{5.46}
\ee
The scalar product is
\be
\langle \Psi \,\Psi \rangle _S  
= \,\psi ^{(x)} \rho _{\,(x)\,*(x')} \psi ^{*(x')} 
 + \psi ^{*(x)} \rho _{\,*(x)\,\,(x')} \psi ^{\,(x')}  .
\lbl{5.47}
\ee

Let us introduce the object
\be
\Omega  = \prod\limits_x {h_{\,*(x)} } ,
\lbl{5.49}
\ee
satisfying
\be
h_{\,*(x)} \Omega  = 0 .
\lbl{5.50}
\ee
We see that $\Omega$ has the role of vacuum, and $h_{\,*(x)}$ are
{\it annihilation operators}. The object $h_{\,(x)}$ gives a state
$h_{\,(x)} \Omega \neq 0$, with a hole in vacuum. Therefore, $h_{\,(x)}$
are {\it creation operators} (of holes in $\Omega$).

If we act with a vector $\Psi$ on the vacuum $\Omega$, we obtain
\be
\Psi \,\Omega \, = \,\psi ^{\,(x)} \,h_{\,\,(x)} \,\Omega .
\lbl{5.51}
\ee
This is a superposition of one particle (or better, one hole) states.

A generalization of eq.\,(\ref{5.51}) is
\be
  (\,\psi _0  + \psi ^{\,(x)} \,h_{\,\,(x)} 
 + \psi ^{(x)(x')} h_{\,(x)} h_{\,(x')}  + \,\,...\,\,)\Omega .
\lbl{5.52}
\ee
This is a superposition of zero, one two,...,many,..., particle states.
Such a state is the infinite dimensional analog\,\ci{PavsicPhaseSpace} of the
spinor as an element of a left ideal\footnote{In the case of a finite dimensional
Clifford algebra $Cl(p,q)$, a left ideal is a subspace of $Cl(p,q)$ that is invariant
under the left multiplication by any element of $Cl(p,q)$.}
 of Clifford
algebra\,\ci{SpinorBasis, PavsicPhaseSpace}.

Besides (\ref{5.50}), there are other possible vacuums, e.g.,
\bear
&&\Omega  = \prod\limits_x {h_{\,(x)} \,,\,\,\,\,\,\,\,\,\,\,\,\
,\,\,\,\,\,\,\,\,\,\,\,} h_{\,(x)} \Omega  = 0 \lbl{5.53}\\
&&\Omega  = \left( {\prod\limits_{x \in R_1 } {h_{\,*(x)} } } \right)
\left( \prod\limits_{x \in R_2 } h_{\,(x)}  \right)~,~~~~~
    h_{*(x)} \Omega = 0,~~{\rm if}~ x \in R_1 \nonumber\\
    &&\hs{6.5cm} h_{\,(x)} \Omega = 0,~~{\rm if}~ x\in R_2 \lbl{5.54}\\
    &&{\rm etc.} \nonumber
\ear
where\footnote{If we consider the Stueckelberg theory\,\ci{Stueckelberg},
then instead of ${\mathbb R}^3$ we have ${\mathbb R}^{1,3}$.}
$R_1 \in {\mathbb R}^3,~R_2 \in {\mathbb R}^3,~R_1 \cup R_2 = {\mathbb R}^3$.

An analogous situation also holds in momentum representation, where the operators are
$c_{\,(p)}$, $c_{*(p)}$, with $p=(p^0,{\bf p})$. The vacuum can then be
defined as, e.g.,
\be
\Omega = \left ( \prod_{p^0>0,{\bf p}} c_{*(p^0,{\bf p})} \right )
   \left ( \prod_{p^0<0,{\bf p}} c_{\,(p^0,{\bf p})} \right ) .
\lbl{5.52a}
\ee
Instead of the above notation,
adapted to the Clifford algebraic description, we will now again use
our notation of sec. 5.2.2,  which is a straightforward generalization of the
usual notation. We have the following operators in momentum space:
\bear
 &&b_s^{\,\bar i} ({\bf p})\,,\,\,\,\,\,\,\,\,\,d_s^{\,\bar i} ({\bf p})
 \lbl{5.55a} \\ 
 &&b_s^{\,{\ul i}} ({\bf p})\,,\,\,\,\,\,\,\,\,d_s^{\,{\ul i}} 
 ({\bf p}) \lbl{5.55b}
\ear
which annihilate the vacuum
\be
\Omega \,\, = \,\,\left( {\prod\limits_{s,{\bf{p}}} {b_s^{\bar i} 
({\bf p})} } \right)\,\left( {\prod\limits_{s,{\bf{p}}} {d_s^{\,\bar i}
 ({\bf p})} } \right)\,\left( {\prod\limits_{s,{\bf{p}}} {b_s^{\,{\ul i}} 
 ({\bf p})} } \right)\,\left( {\prod\limits_{s,{\bf{p}}} {d_s^{\,{\ul i}} 
 ({\bf p})} } \right) .
\lbl{5.55}
\ee
The Fock space states are
\bear
&&b_s^{{\bar i}\dag }({\bf p}) \,\Omega \, , ~~~
\,d_s^{\,{\bar i}\dag }({\bf p}) 
\,\Omega \,,~~~
b_s^{{\,\ul i}\dg} \Omega({\bf p}) \,, ~~~
d_s^{{\,\ul i}\dg} \Omega({\bf p}) ~, ...,
~{\rm and ~all ~many~ particle ~states} \nonumber \\
&&~~~^{\rm positive~ energies} 
~~~~~~~~~~~~~~~~ ^{\rm negative ~energies} \lbl{5.56}
\ear
We see that, though considered nowadays as obsolete, the concept of the
Dirac sea finds its revival with the Clifford algebra approach to field
theory. Besides (\ref{5.55}), other vacua can also be constructed by taking
various combinations of the operators $b_s^{i \dg}$, $b_s^i$, $d_s^{i \dg}$,
$d_s^i$  (see ref.\,\ci{PavsicPhaseSpace}).

In the absence of interactions, the vacuum (\ref{5.55}) has vanishing energy,
i.e., the expectation value of the Hamiltonian (\ref{4.23}) is 
$\langle \Omega^\ddg H \Omega \rangle_S = 0$. In the presence of interactions
that mix the positive and negative energy states, the vacuum $\Omega$ decays
into a superposition of positive and negative energy states. The finite state
with infinitely many positive and negative energy particles,
$|p_1 p_2 ... p_\infty \rangle$, is the state in which all the operators
were removed from the vacuum $\Omega$:
\be
\Psi (t) = b_{s_1 }^{\bar i_1 \dag } ({\bf p}_1 )b_{s_2 }^{\bar i_2 \dag } 
({\bf p}_2 )\,...\,d_{s_1 }^{\bar i_1 \dag } ({\bf p} _1 )
d_{s_2 }^{\bar i_2 \dag } 
({\bf p} _2 )\,...\,d_{s_1 }^{\ul i_1 \dg} (p_1 )d_{s_2 }^{\ul i_2 \dg} ({\bf p}_2 )
\,...\,\Omega \, = \,1
\lbl{5.57}
\ee
The expectation value of the free Hamiltonian (\ref{4.23}) in the latter
state is also zero. The state (\ref{5.57}) also is ``unstable" and can
evolve into another state that is a superposition of the basis states
\be
b_s^{i\dag }({\bf p}) \,,\,\,\,\,\,\,\,\,d_s^{i \dag } ({\bf p})
\,,\,\,\,\,\,\,\,\,\,b_s^{i}({\bf p}) ,\,\,\,\,\,\,\,\,\,d_s^{i}({\bf p}) ~,~~~~
{\rm and~all~many~operator~states}
\ee

In a finite (closed) universe, with a minimal (e.g., Planck) distance,
there would be a maximal absolute value of the particle's energy, and there
would be a finite number of discrete momenta values. The ${\bf p}$ in
eqs.\,(\ref{5.55}) would be discrete. The unstable vacuum $\Omega$ would then decay,
in a finite time, into a finite number of positive and negative energy states.
The latter state would then ``decay" back into $\Omega$ (just as in the
simple model of the oscillator illustrated in Fig.\,9). So there would be
oscillations between $\Psi =\Omega$ and $\Psi =1$. Such would be the
situation for an external observer, ${\cal O}_{\rm ext}$, who does not
disturb the state $\Psi (t)$. But for
an observer who is a part of the universe described by the state
$\Psi(t)$, and who ``nearly" continuously watches (at least a part of)
his universe, the evolution of $\Psi (t)$ is ``altered" according to
the watchdog effect of QM; it is no longer an undisturbed evolution.
The discussion at the end of Sec.\,6.3.1 is valid for fermionic fields as well.

\section{Discussion}

Clifford algebras and Clifford spaces appear as a promising framework
for the unification of particles and (gauge) fields. Such spaces do not
have Lorentzian signature, $(+ - - - - ...)$, but a more general signature
$(+++...---...)$. The Clifford space, $C$, whose tangent space at any of
its point is the Clifford algebra $Cl(1,3)$, has the neutral signature
$(8,8)$. We have investigated the stability of the classical and quantum
harmonic oscillator in the space $M_{1,1}$, which is a subspace of $C$.
It is known\,\ci{Woodard,PavsicPseudoHarm} that the negative energies occurring
in a space with neutral signature pose no problem, if there are no interactions.
We have shown that even the presence of an interaction causing transitions
between positive and negative energy states, need not be problematic, and that
such a system may have stable solutions.

In the case of {\it the classical oscillator}, we have found that certain
interactions prevent the runaway solutions, and make the system
stable. Collisions of an otherwise unstable pseudo Euclidean oscillator
with surrounding particles also stabilize the oscillator. After such
collisions, the particles gain kinetic energy. A material made of such
oscillators would thus increase the temperature of the surrounding
medium, after being immersed into it.
This is very hypothetical, but the history of physics teaches us that we
can never be sure about what surprises are waiting ahead of us. Recall
that quasicrystals, with crystallographically forbidden symmetries\footnote{
Schechtman's discovery of quasicrystals\,\ci{Shechtman} was ridiculed,
because, in view of the established crystalographic theory, it was considered
as impossible. Fortunately, that was a simple experiment and it was not
difficult for other labs to repeat it.}, cannot be explained in terms of
local interactions in three dimensions. They can be explained as regular
crystals in 6-dimensions, projected in 3-dimensions. The Clifford space,
associated with objects in spacetime, has sixteen dimensions, and one can
envisage that there exist in $C$ the crystals that, from the 3D point of view,
appear as quasicrystals. If such an explanation of quasicrystals eventually
turns out to be correct, then the next step would be to investigate
whether the pseudo Euclidean oscillators, allowed by the physics
in Clifford space, actually exist in nature.

In the case of {\it the quantum oscillator} in $M_{1,1}$, we have calculated
numerical solutions for a truncated system. Because the truncated system
is finite, the solution is oscillating, and does not run away into infinity.
For practical reasons we considered the energy modes up to $m,n=4$, so that
all higher energies were cut off. We can envisage a cutoff at much higher
energies. This corresponds to a small minimum cutoff distance\footnote{
It is generally believed that the Planck distance is the minimal distance,
therefore the idea that space or Clifford space is in fact a lattice
makes sense.}
in the space $M_{1,1}$. Then, instead of continuous space $M_{1,1}$,
we have a finite periodic lattice. An analogous situation occurs if we consider an
oscillator in 16D Clifford space.

After having investigated how the theory works on the examples of the
classical and quantum pseudo Euclidean oscillator, we considered quantum
field theories. If the metric of field space is neutral, then there occur
positive and negative energy states. An interaction causes transitions
between those states. A vacuum therefore decays into a superposition of
states with positive and negative energies. Because of the vast phase space
of infinitely many particles, such a vacuum decay is instantaneous---for
an external observer. But we, as a part of our universe, are not external
observers. We are internal observers entangled with the ``wave function''
of the rest of our universe. According to the Everett interpretation of quantum
mechanics, we find ourselves in one of the branches of the universal
wave function. In the scenario with decaying vacuum, our branch can consist
of a finite number of particles. Once being in such a branch, it is improbable
that at the next moment we will find ourselves in a branch with infinitely
many particles. For us, because of the ``watchdog effect" of quantum mechanics,
the evolution of the universal wave function relative to us is frozen to
the extent that instantaneous vacuum decay is not possible. The
Everett interpretation is gaining increasing support among quantum
cosmologists\,\ci{Many worlds}.

We have also pointed out that the field potential $V(\varphi)$ need not
be constant during the evolution of the universe. It may change, and thus
alter a system's dynamics from an unstable to a stable regime.

If there exists a minimal distance (e.g., at the Planck scale),
and if the universe is closed, then it contains a finite, albeit very large,
number of points; it is a sort of lattice. Such a
finite system cannot run into infinity, because there is neither infinite
energy or infinitely small distance, neither infinite number of particles.
Therefore, such a system is stable even in the presence of negative energies.
It  oscillates with large, but finite, amplitudes. In Sec.\,6.3  we showed
how does oscillate a simple quantum finite oscillator. An analogous situation holds
for quantum fields on a finite lattice.

In a field theory with neutral signature, there is an outburst of positive
and negative energy particles, like an explosion, that is eventually
stabilized. This is reminiscent of the Big Bang. Our universe indeed emerged
in an explosion. But in our universe we do not ``see'' equal number of positive
and negative energy particles. Can then such a quasi unstable vacuum
be an explanation for Big Bang?

Description of our universe requires fermions and accompanying gauge fields,
including gravitation. According to the Clifford algebra generalized
Dirac equation---Dirac-K\"ahler equation---there are four sorts of the
4-component spinors, with energy signs a shown in eq.\,(\ref{5.39}). The vacuum of
such field has vanishing energy and evolves into a superposition of positive
and negative energy fermions, so that the total energy is conserved. 
A possible scenario is that the
branch of the superposition in which we find ourselves, has the sea of
negative energy states of the first and the second, and the sea positive
energy states of the third and forth minimal left ideal of $Cl(1,3)$.
According to ref.\,\ci{PavsicSpinorInverse}, the former states are
associated with the
familiar, weakly interacting particles, whereas the latter states are
associated with mirror particles, coupled to mirror gauge fields, and thus
invisible to us. According to the field theory based on the Dirac-K\"ahler
equation, the unstable vacuum could be an explanation for Big Bang.

Introduction of spaces with neutral signature that imply the existence
of negative energies, is potentially significant for further important
progress of theoretical physics. The usual arguments---mostly related to
stability---why such spaces are not suitable for physics, can be
circumvented along the lines indicated in this paper.

\vs{3mm}

\centerline{\bf Acknowledgment}

\vs{1mm}

This research was supported by the Slovenian Research Agency.

\end{document}